\documentclass[12pt,preprint]{aastex}   
\usepackage{natbib} 
\usepackage{emulateapj5}

\def\aap{A\&A} 
\def\apj{ApJ} 
 
\def\apjl{ApJL} 
\def\mnras{MNRAS}

\def\aaps{A\&A Supp.} 
 
\def\apjs{ApJS} 
\def\pasj{PASJ} 
\def\gca{Geochimica et Cosmochimica Acta}

\newcommand{\hm}{\,h^{-1}{\rm Mpc}}  
  
\newcommand{\chandrasp}{{\it Chandra} } 
 
\newcommand{\asca}{{\it ASCA} }

\newcommand{\xmm}{{\it XMM-Newton}}  
\newcommand{\xmmsp}{{\it XMM-Newton} }  
\newcommand{\suzaku}{{\it Suzaku} }  
\newcommand{\xmassp}{{\small X-MAS} }
\newcommand{\xmas}{{\small X-MAS}}
\newcommand{\mekal}{{\small MEKAL} } 
\newcommand{\vmekal}{{\small VMEKAL} } 
\newcommand{\apec}{{\small APEC} } 
\newcommand{\vapec}{{\small VAPEC} } 
\newcommand{\xspec}{{\small XSPEC} } 
\newcommand{\wabs}{{\small WABS} } 
\newcommand{\be}{\begin{equation}} 
\newcommand{\ee}{\end{equation}} 
\newcommand{\ba}{\begin{eqnarray}} 
\newcommand{\ea}{\end{eqnarray}}

\newcommand{\sph}{{\small SPH} }

\shorttitle{\xmas 2: Metallicity Systematics}
\shortauthors{Rasia et al.}

\begin{document}   

\title{X-MAS2: Study Systematics \\ on the ICM Metallicity Measurements}  
\author{
E. Rasia\altaffilmark{1,2,3}, 
P. Mazzotta\altaffilmark{4,5},  
H. Bourdin\altaffilmark{4}, 
S. Borgani\altaffilmark{6,7,8},\\~\\  
L. Tornatore\altaffilmark{9}, 
S. Ettori\altaffilmark{10}, 
K. Dolag\altaffilmark{11},  
L. Moscardini\altaffilmark{12}
}

\altaffiltext{1}{ Department of Physics, University of Michigan, 450 Church St., Ann Arbor, MI 48109-1120, USA }
\email{rasia@umich.edu}
\altaffiltext{2}{ Dipartimento di Astronomia, Universit\`a di Bologna, 
via Ranzani 1, I-40127, Bologna, Italy }
\altaffiltext{3}{ Chandra Fellow }

\altaffiltext{4}{ Dipartimento di Fisica, Universit\`a di Roma Tor Vergata, 
via della Ricerca Scientifica 1,\\ I-00133, Roma, Italy }
\altaffiltext{5}{ Harvard-Smithsonian Centre for Astrophysics, 60 Garden Street, Cambridge, MA 02138, USA}
\altaffiltext{6}{ Dipartimento di Astronomia, Universit\`a di Trieste, via 
  Tiepolo 11, I-34131, Trieste, Italy }
\altaffiltext{7}{ INAF --  Osservatorio Astronomico di Trieste, via Tiepolo 11, I-34131, Trieste, Italy}
\altaffiltext{8}{ INFN -- National Institute for Nuclear Physics, I-34100, Trieste, Italy}
\altaffiltext{9}{ SISSA/International School for Advanced Studies, via Beirut 4, I-34100, Trieste, Italy }
\altaffiltext{10}{ INAF, Osservatorio Astronomico di Bologna, via Ranzani 1, I-40127, Bologna, Italy }
\altaffiltext{11}{ Max-Planck-Institut f\"ur Astrophysik, Karl-Schwarzschild Strasse 1, D-85748,\\ Garching bei M\"unchen, Germany}
\altaffiltext{12}{ INFN, Sezione di Bologna, viale Berti Pichat 6/2, I-40127 Bologna, Italy }

%---------- 

\begin{abstract}   
The X-ray measurements of the intra-cluster medium metallicity are
becoming more and more frequent due to the availability of powerful
X-ray telescope with excellent spatial and spectral resolutions. The
information which can be extracted from the measurements of the
$\alpha$--elements, like Oxygen, Magnesium and Silicon with respect to
the Iron abundance is extremely important to better understand the
stellar formation and its evolutionary history.  In this paper we
investigate possible source of bias or systematic effects connected to
the plasma physics when recovering metal abundances from X-ray
spectra. To do this we analyze 6 simulated galaxy clusters processed
through the new version of our X-ray MAp Simulator (\xmas), which
allows to create mock \xmmsp EPIC MOS1 and MOS2 observations. By
comparing the spectroscopic results inferred by the X-ray spectra to
the expected values directly obtained from the original simulation we
find that: $i)$ the Iron is recovered with high accuracy for both hot
($T> 3$ keV) and cold ($T<2$ keV) systems; at intermediate
temperatures, however, we find a systematic overestimate which
depends inversely on the number counts; $ii)$ Oxygen is well
recovered in cold clusters, while in hot systems the X-ray measurement
may overestimate the true value by a factor up to 2-3; $iii)$ Being a
weak line, the measurement of Magnesium is always difficult; despite
of this, for cold systems (i.e. with $T< 2$ keV) we do not find any
systematic behavior, while for very hot systems (i.e. with $T >5$ keV)
the spectroscopic measurement may strongly overestimate the true value
up to a factor of 4; $iv)$ Silicon is well recovered for all the
clusters in our sample.  We investigate in detail the nature of the
systematic effects and biases found performing \xspec
simulations.  We conclude that they are mainly connected with the
multi-temperature nature of the projected observed spectra and to the
intrinsic limitation of the \xmm-EPIC spectral resolution that does
not always allow to disentangle among the emission lines produced by
different elements.
\end{abstract}
   
%-------------- 
%---------------------------------------------------------   
 
\section{INTRODUCTION}   
\label{sec:intro} 
Measuring the metallicity of the intra--cluster medium (ICM)
represents a unique means towards a unified description of the
thermodynamical properties of the diffuse gas and of the past history
of star formation within the population of cluster galaxies
\citep[e.g., ][]{renzini.etal.93}. The total mass in metals is
directly related to the number of supernovae (SNe) which exploded in
the past. Furthermore, measuring the relative abundance of Iron with
respect to $\alpha$--elements provides information on the relative
fraction of type II and type Ia SN and, therefore, on the shape of the
initial mass function (IMF) for star formation. Finally, the redshift
evolution of the metal content of the ICM reflects the rate at which
metals are released in the diffuse medium. This, in turn is directly
related to the star formation history.

From an observational point of view, measurements of the global
content of different metal species within the ICM have been first made
possible with the advent of the \asca satellite \citep[e.g.,
][]{fukazawa.etal.94,mushotzky.etal.96, fukazawa.etal.98,
  finoguenov.etal.99, finoguenov&ponman99, fukazawa.etal.00,
  finoguenov.etal.00, finoguenov&jones00, matsushita.etal.00,
  finoguenov.etal.01a}. Based on a collection of ASCA data for a
fairly large number of clusters, \cite{baumgartner.etal.05} determined
the relation between the global ICM content of different metal species
and the temperature. At the same time, data from the Beppo--SAX
satellite allowed to perform a detailed study of the spatial
distribution of Iron \citep{degrandi&molendi01}. This study showed that
significant negative metallicity gradients are present in ``cooling
flow'' clusters, with a central metallicity spike associated to the
brightest cluster galaxy \citep{degrandi.etal.04}. More
recently, the excellent spatial resolution of \chandrasp has allowed to
obtain Iron metallicity profiles
\citep{ettori.etal.02b,buote.etal.03,dupke&white03,humphery&buote06}
and maps \citep{sanders.etal.04, sanders&fabian06} for a significant
number of clusters. Furthermore, the large collecting area of \xmmsp is
now allowing to study in detail the abundances of several elements,
thereby providing a test-bed for different models of SNe yields
\citep[][ and reference
therein]{deplaa.etal.06,deplaa.etal.07,boehringer.etal.05}, and to
largely improve the description of the metal content associated to
cluster cool cores \citep{molendi&gastaldello01, Tamura.etal.01,
boehringer.etal.02, Finoguenov.etal.02, Gastaldello&Molendi02,
Sakelliou.etal.02, Matsushita.etal.03,Boehringer.etal.04,
Tamura.etal.04, werner.etal.06a,matsushita.etal.07}. \chandrasp and
\xmmsp allowed to analyze in detail the Iron bias already noticed with
\asca \citep{buote00a}. This bias is arising when fitting with a single
temperature model a plasma characterized by a temperature gradient or
by a multi-temperature structure with the colder component below 1 keV
\citep{molendi&gastaldello01,buote.etal.03}. More recently, the
\suzaku telescope is providing accurate detections of Oxygen and
Magnesium lines and has already provided measurements of the
distribution of $\alpha$--elements in the Fornax cluster
\citep{matsushita.etal.06} and in Abell 1060 \citep{sato.etal.07} out
to fairly large cluster-centric distances. Finally, archival studies
based on \chandrasp and \xmmsp observations are now providing
information on the evolution of the global Iron metal content out to
$z\sim 1.3$ and show a decrease of about 50 per cent of the ICM
metallicity from the present time to that redshift
\citep[][]{balestra.etal.07,maughan.etal.07}.

From the theoretical point of view, a number of methods have been
developed to describe the ICM enrichment associated with the
cosmological hierarchical build up of galaxy clusters.
\cite{delucia.etal.04}, \cite{nagashima.etal.05} and
\cite{bertone.etal.07} studied the ICM enrichment by coupling $N-$body
simulations to semi-analytical models (SAMs) of galaxy formation. This
approach, although quite flexible to explore the space of parameters
relevant for star formation and galaxy evolution, does not provide
detailed information on the spatial distribution of metals within
clusters and on the role of gas--dynamical processes (e.g.,
ram--pressure stripping, turbulent diffusion, etc.) in distributing
metals. In order to overcome these limitations, \cite{cora06}
developed a hybrid technique, which is based on applying SAMs to
non--radiative hydrodynamical simulations. Another approach is that of
simulating clusters with hydrodynamical codes, which included a
treatment of specific effects, such as galactic winds
\citep{schindler.etal.05}, ram pressure stripping
\citep{domainko.etal.06} and active galactic nuclei
\citep{moll.etal.07}, although coupled with simplified descriptions of
star formation. Finally, cosmological hydrodynamical simulations of
galaxy clusters have been presented in the last few years, in which
the production of metals is self--consistently related to the process
of gas cooling and star formation, by accounting in detail for the
contribution from different stellar populations and the corresponding
life--times
\citep{lia.etal.02,valdarnini03,tornatore.etal.04,romeo.etal.06,tornatore.etal.07}.

These numerical descriptions of the ICM enrichment have reached a good
enough accuracy, in terms of physical processes included and
resolution, so as to allow a detailed comparison with observational data
to be performed. It is however clear that for this comparison to be
meaningful, one has to be sure to understand observational biases,
related to the instrumental response of the X--ray detectors onboard
of the different satellites. A typical example of such biases in the
study of the ICM is represented by the measure of the X--ray
temperature. \cite{tsl} showed that the combination of a complex
thermal structure of the ICM and of the finite energy band, within
which X--ray spectra are measured, can bias low the value of the
spectroscopic temperature \citep[see also
][]{vikh06}. Therefore, it is worth asking whether similar
biases also affect the determination of the ICM metallicity through
the fitting of the X--ray spectrum with single-temperature and
single-metallicity plasma models.
 
In this paper we address the issue of the biases in the X--ray
measurement of the ICM metallicity by performing mock observations of
hydrodynamical simulations of galaxy clusters, which include a
detailed description of the process of chemical enrichment and follow
the production of different chemical species
\citep{tornatore.etal.07}.  To this purpose, we generate \xmmsp
observations of both EPIC-MOS1 and EPIC-MOS2 cameras, performed with a
new version of the X-ray MAp Simulator \citep[\xmas,][]{xmas} package,
which now also includes the contribution from metal lines in the
generation of the X--ray spectra.  Taking advantage of this tool, we
can carefully check for the presence of systematics in the X-ray
measurement of the metallicity, also by distinguishing the behavior of
Iron and of $\alpha$--elements, when observations are performed at the
spectral resolution of \xmm. In particularly we investigated the
biases related only to the physics of the plasma. The biases related
to the uncertainties on the calibration and/or on the background
and/or on plasma model are out of the scope of this paper. As such,
this paper mainly focusses on quantifying the possible presence of
these systematics and understand their relation with the complex
thermal and chemical structure of the ICM.  We defer to a forthcoming
analysis a comparison between results of hydrodynamical simulations of
ICM enrichment and observational data (Fabjan et al. in preparation).
 
The plan of the paper is as follows.  The new version of our X-ray MAp
Simulator, X-MAS2, will be presented in Sect.~\ref{sec:xmas2}. The
characteristics of the hydrodynamical simulations are described in
Sect.~\ref{sec:sim}. The procedure followed for the X-ray analysis is
presented in Sect.~\ref{sec:xray}. Our results are described in
Sect.~\ref{sec:results}, where we also provide a critical discussion
about the effect of observing X-ray spectra with the typical \xmmsp
energy resolution (Sect.~\ref{sec:basic}). These results are discussed
in Sect.~\ref{sec:discussion}, while the main conclusions of our
analysis are summarized in Sect.~\ref{sec:conclu}. 

All errors quoted are at 1 $\sigma$ level (68.3 per cent level of
confidence for one interesting parameter).

%%%%%%%%%%%%%%%%%%%-----------------------------%%%%%%%%%%%%%%%%%%% 
\section{\xmas 2} 
\label{sec:xmas2} 
 
The numerical code \xmassp \citep{xmas} produces mock \chandrasp and
\xmmsp observations starting from outputs of hydro-N-body simulations
of galaxy clusters.  The software is composed by two main parts: the
first section computes the emissivity of each gas particle contained
in the selected field of view and projects it into the sky plane; the
second one convolutes the flux reaching the telescope using the
response of a given detector.  One of the main characteristics of
\xmassp is that it is able to create the events file containing all
the spectral information for each ``incoming simulated
photon''. Moreover it properly reproduces the spatial features and the
characteristics of the \chandrasp and \xmmsp detectors.  Furthermore
the header of the final event file stores all the fundamental keywords
common to true observed files.  In this way the application of the
routines and analysis programs commonly used by observers is
straightforward. In previous work the code has been used only in the
\chandrasp ACIS-S3 mode to address several issues, like the difference
between the emission-weighted and spectroscopic temperature
definitions
\citep{xmas}, the validity of the spectroscopic-like formula proposed 
by \cite{tsl} and the possible sources of bias in the X-ray estimates 
of cluster masses \citep{rasia.etal.05,rasia.etal.06}.  To do that the 
ICM metallicity  has been kept fixed at 0.3 solar 
\citep[assuming][]{anders&grevesse} and the emissivity has been computed 
simply adopting a MEKAL model
\citep{mewe.etal.85,mewe.etal.86,liedahl.etal.95}. For the purpose of 
this paper a much more detailed analysis of the metal content of the
ICM is required. Therefore, \xmassp has been subject to a substantial
upgrading which affected both the first and second parts.
 
\subsection{First part: implementing the metallicity}
 
We refer to \cite{xmas} for a detailed derivation of the incoming 
photons flux, $F^{\gamma}_{\nu}$. Here we report its final expression 
as a function of redshift $z$, luminosity distance $d_L$, Emission 
Measure $EM \equiv \int_V n_e n_H dV'$ (where the integrals of the 
electron, $n_e$, and hydrogen, $n_H$, densities are computed over the 
volume $V$) and power coefficient $P^{\gamma}_{\nu(1+z)}(T,Z)$: 
\be 
F^{\gamma}_{\nu}= \frac{(1+z)^2}{4\pi d^2_L} EM P^{\gamma}_{\nu(1+z)}(T,Z)\ . 
\ee 
Notice that the power coefficient is a function of both temperature and
metallicity. 
To take this latter into account, the metal content is split in different
elements and their  contribution is
computed separately.
 
Thus, the total power coefficient per each particle  is given by: 
\begin{eqnarray} 
\label{eq:pow} 
P^{\gamma}_{\nu(1+z)}(T,Z_{\rm tot})  &=&  P^{\gamma}_{\nu(1+z)}(T,H) + \\ 
&&\Sigma_i  m(Z_i)[P^{\gamma}_{\nu(1+z)}(T,Z_i) - P^{\gamma}_{\nu(1+z)}(T,H)]\ , 
\nonumber 
\end{eqnarray} 
where the sum is over each chemical element, $Z_i$, considered in the
simulation (He included). The power coefficients are computed assuming
a single-temperature \vmekal model, where we fix to 1 solar value the
contribution of each element, namely He, C, N, O, Mg, Si and Fe. The
spectrum computed by \vmekal for each element,
$P^{\gamma}_{\nu(1+z)}(T,Z_i)$, will be the sum of the continuum,
given by the power coefficient corresponding to hydrogen,
$P^{\gamma}_{\nu(1+z)}(T,H)$, plus the contribution of the different
elements. The factor $m(Z_i)$ represents the weight equal to the
$Z_i$-element mass in units of its Solar values
\citep[defined using the abundances reported in][]{anders&grevesse}.
Hereafter we will compute the emissivity by using a \mekal model,
nevertheless the code includes a flag to switch to the \apec model
\citep[and corresponding \vapec for multi-metal elements treatment, see]
[]{smith.etal.01} whenever desired.
 
 In the new version of the simulator we have also modified the
routine based on the \sph kernel \citep{monaghan&lattanzio} to
distribute the particle emissivity, and assign the interesting
quantities to the projected pixels.  The change consists in computing
the integration over the area through the products of two integrals
along the two directions of the plane, where each integral has been
analytically calculated at priori. This trick significantly speeds up
the computational time.
 
As in the previous version of \xmas, we also include the effect of the
Galactic HI absorption. Chosen a value of the column density $N_H$,
we multiply the flux in each energy channel by an absorption
coefficient computed adopting the \wabs model
\citep{morrison&mccammon}. 
 
\subsection{Second part: simulating EPIC observations}
 
Once the projected flux has been obtained by the first unit of
\xmas, the second unit takes care of the simulation of cluster
``observations''.  This second part has been implemented to simulate \xmmsp
observations made with the EPIC camera. This is done by  
generating a  set of photons, given the
incoming ICM and chosen an exposure time. These are then applied to a
ray-tracing procedure designed to mimic the main characteristics of
the telescope optical paths and detector responses.
\citep[See also][]{bourdin.etal.04}.
 
The incoming set of photons $\{p(k,l,e)\}$, having position on mirrors
$[k,l]$ and individual energy $e$, is generated from random
realizations of the total energy expected assuming a given ICM flux
$F_{\nu}^{\gamma}$, a telescope area $A$, and an exposure time $t$.
At this aim, we project the energies $t \times A \times
F_{\nu}^{\gamma}$ along the line of sight, and we store the obtained
quantity in a three-dimensional array (where two dimensions represent
the position on the mirrors and the third one is for energy).
Assuming a Poisson statistics, we can finally associate a discrete
number of photons to the resulting total energies. Initially we
consider an array having angular and spectral resolutions of $\Delta r
= 3$ arcsec and $\Delta e = 20$ eV,, which are below the expected
performances of the \xmm-EPIC instruments.
 
The ray-tracing procedure emulating the EPIC observations comprises a 
set of filtering and redistribution functions of the photon energies 
and positions across the detector planes. 
Information about the different instrumental 
effects are coming from ground and on-flight calibrations, and 
essentially provided in the \xmmsp Current Calibration Files (CCF) and 
EPIC Redistribution Matrix Files (RMF). The main steps of our 
procedure are as follows: 
 
\begin{itemize} 
\item redistribution of the photon coordinates, $[k,l]$, according to the 
instrument Point Spread Functions (PSF), given as a function of the
position on mirrors and energy of the incidental photons;
\item rejection of part of the incoming photons due to the spatially 
dependent cross-section of mirrors, the so-called ``vignetting''; 
\item suppression of photons falling outside CCDs or within dead 
pixels;
\item rejection of part of the incoming photons due to energy 
dependent effective area of mirrors, filter absorption, and quantum
efficiency of detectors;
\item redistribution of photon energies $e$, according to the detector 
response; 
\item addition of particles induced false detections, with random 
drawing performed following the relative energy spectrum expected for 
a given exposure time. 
\end{itemize} 
 
We finally obtain a list of events, $\{ev(k,l,e)\}$, corresponding to 
photon detections -- but also including some false ones due to 
particles  --, with registered positions on the detector plane, $[k,l]$, 
and energies, $e$. 
 
%%%%%%%%%%%%%%%%%%%-----------------------------%%%%%%%%%%%%%%%%%%% 
\section{THE SIMULATED CLUSTERS} 
\label{sec:sim} 
%---------------------%---------------------%---------------------% 
 
The six simulated clusters analyzed in this paper have been selected
from a larger ensemble of simulations of 19 clusters presented by
\cite{saro.etal.06}; we refer to that paper for more details. These
clusters are extracted from a parent DM-only simulation
\citep{yoshida.etal.01} with a box size of $479\,h^{-1}$Mpc of a flat
$\Lambda$CDM model with $\Omega_{\rm 0m}=0.3$ for the present matter
density parameter, $h=0.7$ for the Hubble constant in units of 100 km
s$^{-1}$Mpc$^{-1}$, $\sigma_8=0.9$ for the r.m.s. fluctuation within a
top-hat sphere of $8\hm$ radius and $\Omega_{\rm 0b}=0.04$ for the
baryon density parameter.  Mass resolution is increased inside the
interesting regions by using the Zoomed Initial Condition (ZIC)
technique proposed by \cite{tormen.etal.97}. Besides the
low-frequency modes, which were taken from the initial conditions of
the parent simulation, the contribution of the newly sampled
high-frequency modes was also added.  Once initial conditions are
created, we split particles in the high-resolution region into a DM
and a gas component, whose mass ratio is set to reproduce the assumed
cosmic baryon fraction. In the high-resolution region, the masses of
the DM and gas particles are set to $m_{\rm DM}=1.13\times
10^9\,h^{-1}{\rm M}_\odot$ and $m_{\rm gas}=1.7\times 10^8\,h^{-1}{\rm
M}_\odot$, respectively. The Plummer-equivalent softening length for
the gravitational force is set to $\epsilon_{\rm Pl}=5\, h^{-1}$kpc,
kept fixed in physical units from $z=5$ to $z=0$, while being
$\epsilon_{\rm Pl}=30\, h^{-1}$kpc in comoving units at higher
redshifts.
 
The simulations have been carried out with a version of the {\small 
GADGET-2} code \footnote{http://www.MPA-Garching.MPG.DE/gadget/} 
\citep{gadget2}, which includes a detailed treatment of chemical 
enrichment from stellar evolution
\citep{tornatore.etal.04,tornatore.etal.07}. {\small GADGET-2} is a
parallel Tree+SPH code with fully adaptive time-stepping, which
includes an integration scheme which explicitly conserves energy and
entropy \citep{springel&hernquist02}, the effect of a uniform and
evolving UV background \citep{haardt&madau}, star formation from a
multiphase interstellar medium and a prescription for galactic winds
triggered by SN explosions (see \citealt{springel&hernquist03} for a
detailed description), and a numerical scheme to suppress artificial
viscosity far from the shock regions 
\citep[see][]{dolag.etal.05a}. In the original version of the code, 
energy feedback and global metallicity were produced only by SNII
under the instantaneous-recycling approximation (IRA).  We have
suitably modified the simulation code, so as to correctly account for
the life-times of different stellar populations, to follow metal
production from both SNIa, SNII, as well as from low and intermediate
mass stars, while self-consistently introducing the dependence of the
cooling function on metallicity \citep{sutherland&dopita}. A detailed
description of the implementation of these algorithms is presented in
detail by \cite{tornatore.etal.07}. The simulations analyzed here assume the 
power-law shape for initial stellar mass function, as proposed by 
\cite{salpeter} and galactic ejects with a speed of 500 km s$^{-1}$.

\begin{figure*}
\begin{center} 
%scale=linear, color=bb 
%MIN 0.1 MAX 13,11,4.5,3 for g1.a, g51, g1.b and all the others 
%REGIONS: thick=2, color=green, properties=include 
%CONTOURS: thick=2, color=black, 10 logaritmic steps between 
%[5e-06 5e-03] for g1.a and g51 
%[5e-07 5e-04] for all the others 
\includegraphics[width=0.8\textwidth]{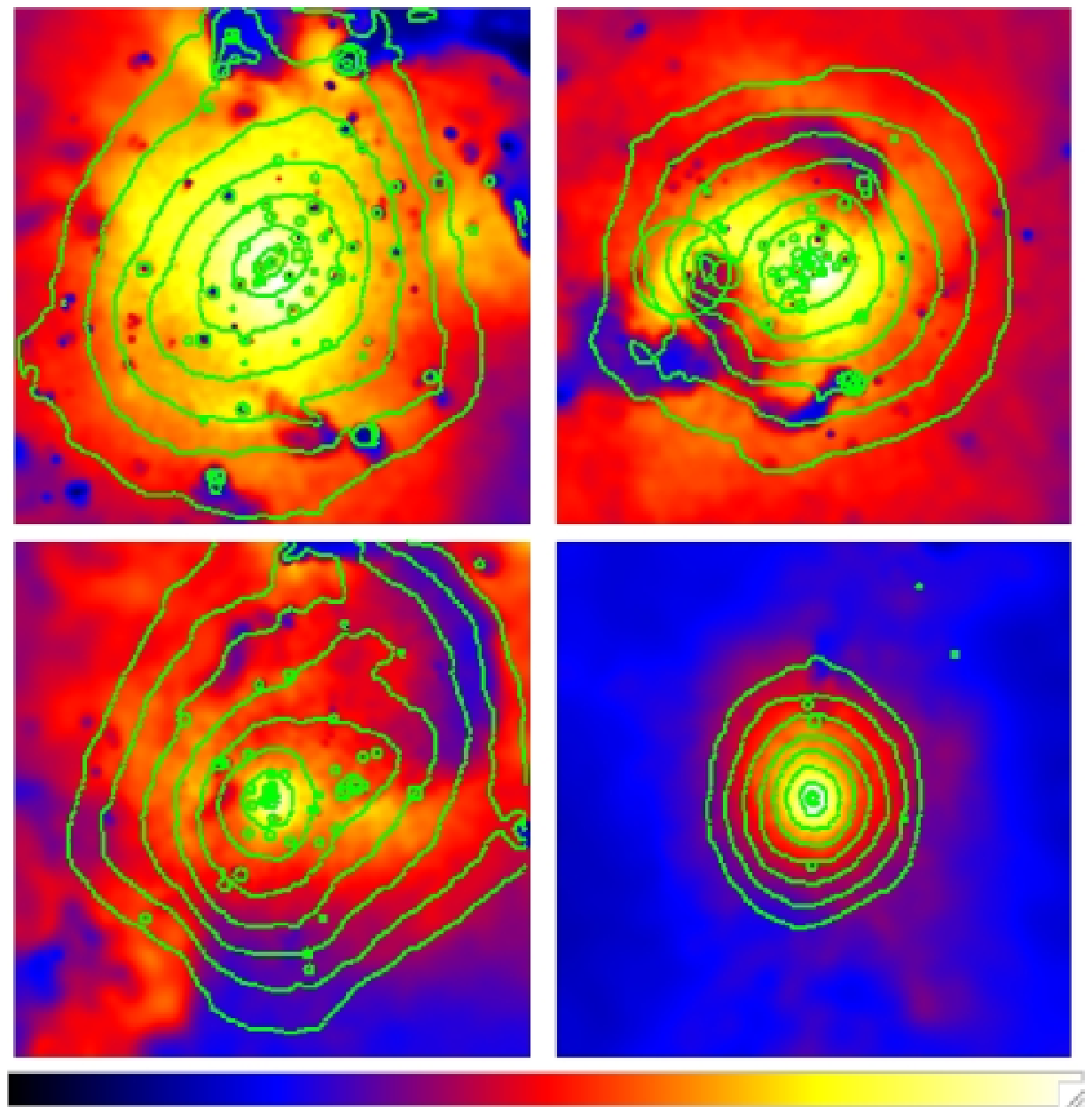}
\caption{Spectroscopic-like temperature maps for the mock clusters: 
G1.a (top-right panel), G51 (top-left panel), G1.b (bottom-right
panel), G676 (bottom-left panel).
The color scale is linear, ranging from a minimum value of 0.1 keV up
to a maximum value, which is equal to 13, 11, 4.5, 3 keV for G1.a,
G51, G1.b and G676, respectively.  We overplot (black curves) the flux
contours (level equi-spaced by a factor of 2) and the compact X-ray
bright cool regions detected using the wavelet decomposition algorithm
(green circles).  }
\label{fig:tsl}
\end{center}  
\end{figure*}

\begin{figure*}
\begin{center}  
%REGIONS: text=helvetica,14,bold 
%SCALE=linear, max and min in the plots 
\includegraphics[width=1\textwidth]{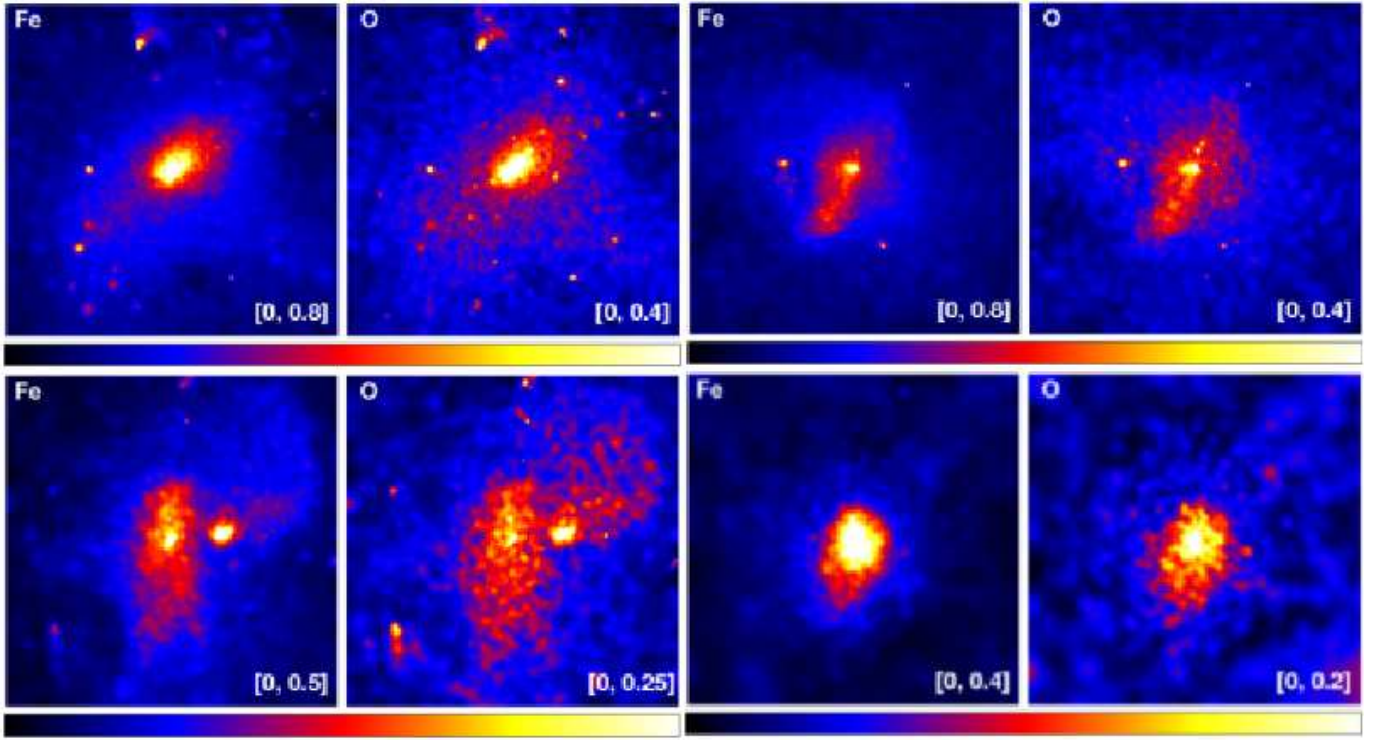}
\caption{Iron and Oxygen distribution map for the same clusters shown in Fig.~\ref{fig:tsl}: G1.a (top-right panel), G51 (top-left panel), G1.b (bottom-right panel), G676 (bottom-left panel).
The linear color scale (in solar units, Anders \& Grevesse) is
reported in each panel.}
\label{fig:metals} 
\end{center} 
\end{figure*}

The main global properties of the simulated clusters are reported in
Table~\ref{tab:clusters}. In the sample there are two big objects with
mass (computed inside the radius $R_{200}$ which comprises an
over-density of 200 over the Universe critical density) larger than $10^{15}
M_{\odot}$, a medium-mass cluster ($M_{200}=3.5 10^{14}M_{\odot}$) and
three smaller systems, namely G676, G914 and G1542. Since these last
three objects have very similar characteristics and behaviors, we
chose to report our results only for the first one, as example.  In
Fig.~\ref{fig:tsl} and Fig.~\ref{fig:metals} we show the projected
maps for the spectroscopic-like temperature and for the metallicity,
respectively.  The map size is equal to the field of view of \xmm,
i.e. 30 arcmin: since we put our clusters at redshift $z=0.06$, this
corresponds to 2088 kpc, using the cosmological model here assumed.
The different metal maps refer to Iron and Oxygen (representative of
the $\alpha$--elements). From these maps it appears that the largest
clusters, G1.a and G51, have a rather complex structure, while the
smaller cluster G676 appears as more relaxed, with G1.b representing
an intermediate case.
 
\begin{figure*}
\begin{center} 
% PHOTON IMAGES OFTHE CLUSTERS obtained with 
%dmcopy "g914.a.008.z_2e5.xmm_m2.fits[pi=700:2000][bin x=64,y=64][part=0,flag=0]" g914.a.008.z_2e5.xmm_m2_PURE.img clob=yes. 
%scale parameter=200 and color bb 
\includegraphics[width=1\textwidth]{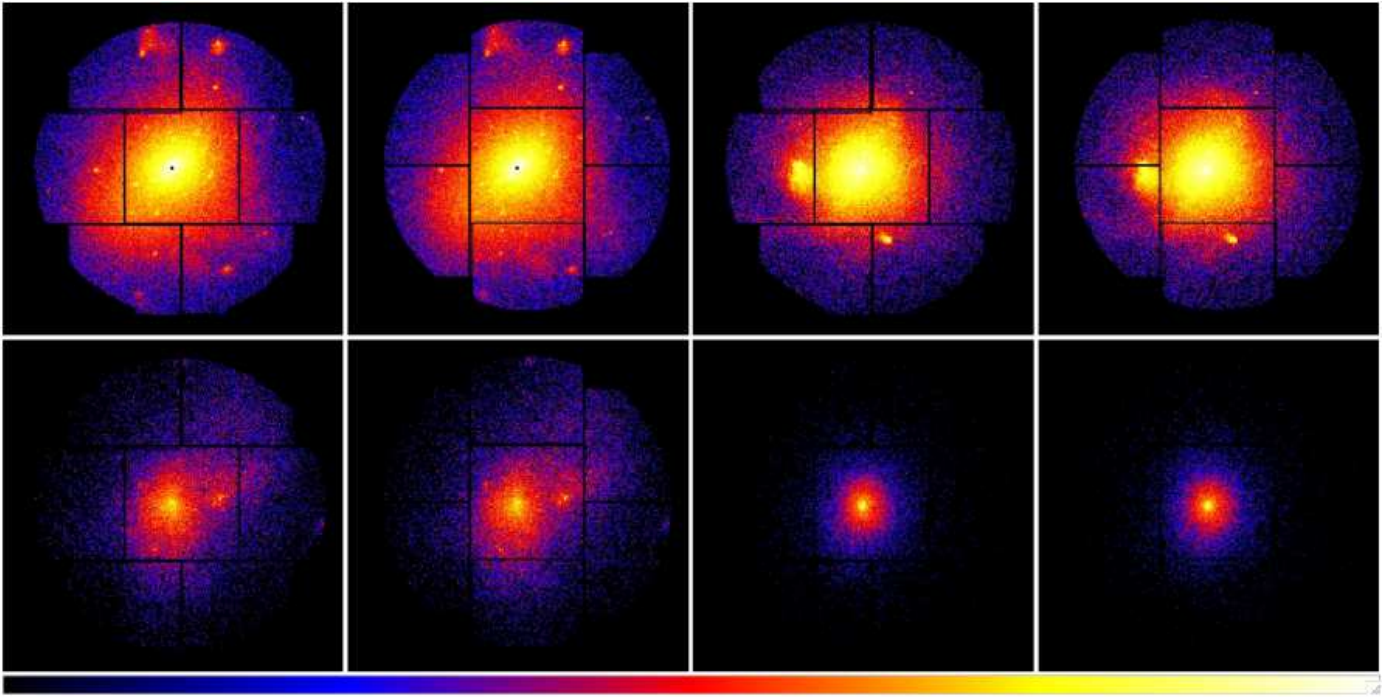}
\caption{Photons images in the [0.7-2] keV energy band of the clusters in  fig~\ref{fig:tsl}: G1.a (top-right panel), G51
(top-left panel), G1.b (bottom-right panel), G676 (bottom-left penal).
The images are  binned to 3.2 arcsec, background subtracted, and
 corrected for vignetting and  out-of-time events}
\label{fig:clusters} 
\end{center}
\end{figure*}

In the following we briefly describe the dynamical state of our
simulated clusters. This is fundamental to explain the sample
characteristics which could influence the presence of any bias.  The
two massive clusters, G1.a and G51, both have an active dynamical
state and strong inhomogeneities in the temperature maps (upper panels
of Fig.~\ref{fig:tsl}).  They show a monotonically increasing mass
history, especially lately: this means that there is a significant
amount of material which continuously in-falls inside the cluster
virial region.  All these small structures can be recognized both in
the temperature maps (Fig.~\ref{fig:tsl}) and in the metallicity maps
(Fig.~\ref{fig:metals}) of these objects (see also the photon images
reported in Fig.~\ref{fig:clusters}). Those are cold blobs which have
not had enough time to thermalize in the hot ICM. Their presence does
not affect the results of our analysis, since they have been
identified and excluded from the X-ray analysis (the masked regions
are shown as green circles in Fig.~\ref{fig:tsl}, see
Sect.~\ref{sec:xray}). The metal maps show rather elongated
structures, which trace the direction of the most recent merger
events.

The cluster G1.b is close to G1.a, lying in the same re-simulated
Lagrangian region. It has experienced two consecutive important major
merger events and it has not had enough time to relax. This clearly appears
from the complex features in the temperature map (see also the presence on
the photon image of a bright extended source on the West side of the
centre). The core of the merged halo is detectable by eye from the
metal maps. Quite interestingly, the Oxygen distribution can be used
to reconstruct the merging dynamics.

The cluster G676 is an isolated system, which presented a relevant
merging event at $z\approx 0.7$, but after then it has had time to
relax. The object can be considered in dynamical equilibrium at
present time, in the sense that its global velocity dispersion and its
mass are almost constant in time after the merging event. Its
homogeneous temperature map, round flux contours and spherical
metallicity distribution confirm this picture.  Similar
considerations can be done for the other two small objects, G914 and
G1542 (not shown in the figures).

%%%%%%%%%%%%%%%%%%%-----------------------------%%%%%%%%%%%%%%%%%%% 
\section{The X-ray analysis} 
\label{sec:xray} 

The simulated clusters have been processed through \xmas 2 in order to
produce mock observations with the EPIC-MOS1 and EPIC-MOS2 cameras and
with an exposure time of 200 ksec. As mentioned before, the field of
view is 30 arcmin, corresponding to 2088 kpc in physical units. The
images have been created by integrating along the line of sight a box
of 10 Mpc centered on the clusters. In our analysis we assume that
\xmmsp is perfectly calibrated.
 
%---------------------%---------------------%---------------------% 
\subsection{Spatial analysis} 
 
We create the photon images in the [0.7-2.0] keV band and use them to
detect the presence of compact X-ray bright gas regions.  These
corresponds to cool and very dense gas clumps present in the simulation
and likely produced by overcooling effects. These clumps are generally
not observed in real clusters so to avoid contamination we remove
these regions from our analysis.  For this purpose we used the wavelet
decomposition algorithm proposed by
\cite{vikh.etal.98}. Parameters are settled to localize sources 
with a peak above 4.5$\sigma$ and to identify the surrounding region above
the 2.5$\sigma$ level. 
The selected areas are shown as green circles in the temperature maps
(Fig.~\ref{fig:tsl}).  In Fig.~\ref{fig:clusters} we show the photons
images, extracted in the [0.7-2] keV energy band, corrected by
vignetting, out-of-time events, point spread function, background and
out-field-of-view component, binned to 3.2 arcsec.

%---------------------%---------------------%---------------------% 
\subsection{Spectral analysis} 
 
The spectral analysis has been performed by extracting the spectra
from the event file in concentric annuli, centered on the brightest
X-ray peak of the cluster. The first annulus has an internal radius of
50 kpc (the central region is excluded from the analysis) and an
external radius which is 1.25 times larger. All the remaining annuli
are logarithmically equi-spaced with the same step of 1.25. During the
spectrum extraction, we also subtracted all photons coming from the
masked regions, already mentioned in the previous section, and from
box regions containing the CCD gaps (clearly detectable as horizontal
and vertical lines in the photon images displayed in
Fig.~\ref{fig:clusters}). The spectra have been fitted
using the C statistics in the \xspec 11.3.2 package \citep{arnaud96}
and following two procedures widely used by observers:
 
\begin{itemize} 
\item[-] Method 1. We fit the data with an absorbed VMEKAL model in the [0.4-8] 
keV energy band, leaving at the same time as free all the parameters
of interest: temperature, Iron, Silicon, Magnesium, Oxygen and
normalization. This approach is faster and more automatic but can lead
to artificially biased results because of the degeneracy of the
parameters.
 
\item[-] Method 2. This is a four steps procedure.
 (i) we fit the data with an absorbed MEKAL model in the [0.4-8] keV band to 
 obtain the temperature; metallicity and normalization are considered free parameters; 
 (ii) we fix the temperature and use a VMEKAL model in the same energy 
 band to recover the Iron abundance; the other three metals are left free; 
 (iii) we keep frozen temperature 
 and Iron to measure the Oxygen in the [0.4-1.5] keV band and (iv) finally 
 we freeze the values of temperature, Iron, Oxygen to 
 estimate the  Magnesium and Silicon abundances in the [1.2 3.2] keV 
 band. 
 
\end{itemize} 
 
In both fitting methods we fixed the Galactic absorption $N_H$ and
the redshift to the input values used to simulate the observations with
 \xmas 2 (i.e. $N_H=5\times
10^{20}$ cm$^{-2}$ and $z=0.06$).
 
%%%%%%%%%%%%%%%%%%%-----------------------------%%%%%%%%%%%%%%%%%%% 

%%%%%%%%%%%%%%%%%%%-----------------------------%%%%%%%%%%%%%%%%%%% 
\section{Results} 
\label{sec:results}

From the spectral analysis we recovered the projected 
profiles for temperature, Iron,
Silicon, Magnesium and Oxygen and we compared them to the values
directly extracted from the hydrodynamical simulations.  The
simulation profiles are computed projecting the quantities along the
line of sight, within the same radial bins used in X-ray analysis and
masking out the same regions described in Sect.~\ref{sec:xray}. The
temperature is computed using the spectroscopic-like definition,
$T_{\rm SL}$ \citep{tsl}, summing  over all particles with temperature higher 
than 0.5 keV. For
the metal profiles we use instead the emission-weighted formula: 
\be 
Z_{\rm ew}=\sum Z_i \times W_i/\sum W_i, 
\ee 
where $W_i$ is the emissivity of the $i$-th particle computed in
the [0.4-8] keV energy band considering the contribution of each particle metals.
 
The results for G1.a, G51, G1.b,
and G676 are presented in Fig.~\ref{fig:prof1}.
The black lines refer to the simulated input values, while
spectroscopic values determined from the full-band-fitting (Method 1)
and the ones recovered in the narrow band are shown by the green and
red lines, respectively. The vertical error bars are at
1$\sigma$ level while horizontal bars represent the width of the
radial bin.
 
%---------------------%---------------------%---------------------% 
\subsection{Temperature} 
 
The recovered spectroscopic temperature matches well the input
temperature leading to a maximum relative error of 5 per cent for hot
systems. This difference is expected for objects with a complex
temperature structure (see Fig.~\ref{fig:tsl}) and steep temperature
gradients (see Fig.~\ref{fig:prof1}), as already delineated in
\cite{tsl}.  It is worth noting that the excellent agreement is also
valid for the three cooler systems, for which the temperature falls
from 2 keV at 2 arcmin ($\sim$ 150 kpc) down to 1 keV at 6 arcmin
($\sim$ 450 kpc). Indeed, this result is expected since these objects
show a regular and homogenous temperature map and any weighting
(i.e., mass-weighted, emission-weighted and spectroscopic-like) give
the same consistent result in the limit of isothermal systems
\citep{tsl}.
 
As expected, we do not find any difference between the spectroscopic
temperatures recovered using Methods 1 and 2, i.e. between \mekal and
\vmekal fitting.
In fact we find that for the most massive clusters the two estimates
almost perfectly coincide; in the other cases the differences are
always smaller than 2 per cent.
 
%---------------------%---------------------%---------------------% 
\subsection{Iron} 
 
Thanks to the powerful emission of both Fe-L and Fe-K lines, the
spectroscopic determination of the Iron abundance is recovered with
great accuracy and presents relatively small error bars. The strength
of Fe-K lines for objects having a temperature larger than 3 keV and
the one of Fe-L lines for systems with temperature lower than 2 keV
(see Fig.~\ref{fig:spec}) ensures an appreciable agreement for both
cold and hot clusters. In each radial bin the Iron abundance derived
spectroscopically matches the emission-weighted value from the
original hydrodynamical simulation at 3$\sigma$ level.  Moreover the
relative difference, [Fe$_{\rm spec}$-Fe$_{\rm sim}$]/Fe$_{\rm sim}$,
is less than 5 per cent. The only exception is for G1.b, the 2-3 keV
cluster (see Fig.~\ref{fig:prof1}).  For this object we detect a
systematic overestimate with respect to the input profile which can be
as high as 20 per cent.  The reason of this discrepancy - related to
the complex temperature structure of this cluster (already evident in
the corresponding panel of Fig.~\ref{fig:tsl}) and to its particular
temperature range - will be extensively discussed later in
Sect.~\ref{sec:iron}.  Some spatial bins of G676 show an underestimate
of Iron, which is between 10\% and 30\%. This feature is caused by the
Iron bias described in \citet{buote00a}.  Among our cold clusters,
G676 is the only one presenting this characteristic; we will further
investigate this aspect in Sect.~\ref{sec:ironbias}.

It is worth noting that the Iron spectroscopic abundances obtained from
Method 1 and Method 2 are not significantly different. In fact, the
two procedures differ just by keeping fixed the temperature in the
second case, but the possible degeneracy between Iron and Temperature
is weak because of the large number of counts we had (larger than
$10^4$). The maximum discrepancy between the two methods is smaller
than 10 per cent and it is present only for some radial bins of the
coolest objects.
 
%---------------------%---------------------%---------------------% 
\subsection{Oxygen} 
 
The spectroscopic measurements of Oxygen, as well as Magnesium and
Silicon, are more uncertain in the inner regions of the clusters.  The
effect is larger for the hottest systems, G1.a and G51, and decreases
for objects with smaller masses. The explanation resides on projection
effects: in fact the spectrum of the central region is obtained by
integrating along the whole line-of-sight, thus it includes
contributions from plasma at temperatures which can significantly
differ in the case of most massive clusters, because of their steeper
radial profile.  For the systems with temperature lower than 3 keV,
the spectroscopic determination of the Oxygen abundance always agrees
with the input values inside 1-2$\sigma$.  On the other hand, for the
largest clusters, G1.a and G51, the Oxygen detection is systematically
overestimated, even if the large error bars reduce the statistical
significance in most radial bins.  Increasing the exposure time,
i.e. the photon number, reduces the error and the systematic effect
becomes more evident.  To confirm this result, in Fig.~\ref{fig:oxi}
we report the spectroscopic result for G1.a, our hottest cluster, for
which we simulated a 1 Msec exposure.  From this plot we  notice
how the spectroscopic values significantly  overestimate the input value 
by a factor of 3.  
 
As a general behavior, we notice that Oxygen obtained by fitting the
spectra with Methods 1 and 2 are similar for the smallest objects and
with discrepancies always inside the error bars.  Nevertheless, there
is a tendency of obtaining an abundances larger than the input one
when Oxygen is measured using the broad band (Method 1 shown as green
dashed points in Fig.~\ref{fig:prof1}), being the difference increasing
with the temperature. 
 
%---------------------%---------------------%---------------------% 
\subsection{Magnesium} 
\label{sec:mag} 
 
\begin{figure*}
\begin{center}  

\includegraphics[width=0.5\textwidth]{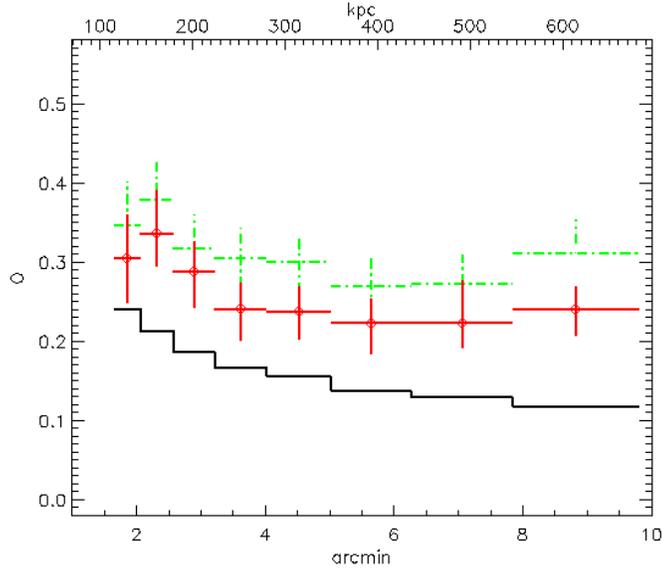}
\caption{
The red lines represent the spectra obtained from \xspec reshifted at
$z=0$ assuming a VMEKAL model with all the metals equal to zero but O,
Si, Mg and Fe which are fixed to the solar values: the temperature is
varying from 1, 2, 3 (top panels starting from the left) to 5, 8, 10
keV (bottom panels).  The separated contributions of the different
elements are also plotted: Oxygen, Magnesium, Silicon and Iron are
shown by green, blue, cyan and black to lines.  All spectra are
convolved with the response of EPIC MOS-1 camera; very similar results
can be obtained considering the EPIC MOS-2 camera. }
\label{fig:spec}
\end{center}   
\end{figure*}

Generally speaking, the measurement of Magnesium is very difficult
because, at the spectral resolution of \xmm,  its line energy 
is well within the energy range of the Fe-L line complex (see
Sect.~\ref{sec:basic} and Fig.~\ref{fig:spec}).
In few cases it is really impossible to detect any Magnesium content, 
and when a detection is obtained we get always very large errors.  
Things improves slightly if we use Method 2 but not by much.
 
Magnesium lines are stronger for cool systems: in fact, in the cases
of G1.b and G676 (see the third and fourth columns of
Fig.~\ref{fig:prof1}), we are able to trace the overall profile, even
if no strong constraint can be obtained.  In any case for these cold
clusters we do not find any systematic bias. On the contrary, the 6
keV cluster, G51, shows large discrepancies in the external bins where
the spectroscopic estimates are between 200 and 400 per cent higher
than the input values (see Fig.~\ref{fig:prof1}). Finally, for G1.a no
conclusion can be reached since its high temperature (all radial bins
have values above 8 keV) makes the Magnesium extremely difficult to be
detected (see also Fig.~\ref{fig:spec}).

%---------------------%---------------------%---------------------% 

\subsection{Silicon} 
 
The difference between the results obtained using Methods 1 and 2 is
significant also in the measurements of the Silicon abundances.
Shrinking the energy band improves the estimates, mainly for
temperature larger than 2-3 keV.  For the cluster G1.b, the choice of
a narrow energy band (red lines in Fig.~\ref{fig:prof1}) helps in
preventing the underestimate of this metal by 40-50 per cent. G51 and
G1.a show a similar tendency, even if this result is less significant
because of the larger error bars. The colder clusters, instead, do not
show any particular difference between the two spectral methods: in
both cases the input abundance of Silicon is well recovered from the
X-ray analysis at 1-1.5 $\sigma$ level.
 
In general the agreement between the X-ray results and the input
values from the hydrodynamical simulations is remarkably good at all
temperatures.  It is important to say, however, that the silicon line
energy is located in region where, due to a combination of CCD and
mirrors, the effective area falls down rapidly and after 2 keV it
shows a strong edge due to Au and Ir. For real observations this may
be a problem as this spectral region of the detector is very difficult
to calibrate.  However, this problem will not be present for
high-redshift clusters, for which the Silicon line is redshifted
towards the softer region.
 
%-------- 
\subsection{Thermal Breemstrahlung spectra as seen by \xmm} 
\label{sec:basic} 
 
Our main goal is to investigate how well we can measure the metal
abundances with the availability of the \xmmsp EPIC camera spectra.
In order to better understand the results of our analysis we
illustrate here some general features of thermal breemstrahlung
spectra observed with the same spectral resolution of \xmmsp
detectors.  We generate with \xspec six spectra at different values of
temperature: 1, 2, 3, 5, 8, and 10 keV. These are convolved with the
instrumental response of the MOS-1 camera (almost identical results
can be obtained considering MOS-2). The adopted plasma model is
VMEKAL, for which we set to zero the abundances of all metals, except
O, Mg, Si and Fe, which are fixed to their solar values.  The
resulting spectra are shown as red lines in Fig.\ref{fig:spec}. In the
same plot different color lines refer to the separate contribution of
each metal: Oxygen (green), Magnesium (blue), Silicon (cyan) and Iron
(black).  Decreasing the plasma temperature, we can notice that the
emission lines become more evident and the ratio between Fe-L (Iron
lines around 1 keV), and Fe-K (Iron lines around 6 keV), rapidly
increases.  The Oxygen lines, due to both OVII and OVIII lines, show
two bumps around 0.6-0.8 keV which are more evident for systems with
temperature lower than 3 keV.  The Oxygen lines are located very close
to the Fe-L complex, whose tail in the soft energy region, given the
spectral resolution of \xmm, can interfere with Oxygen determination
for systems with temperature around 1-2 keV. At temperatures larger
than 3 keV, the Oxygen and Fe-L features are well separated.  The
Magnesium lines are quite strong only for very cold systems with
temperature around 1-3 keV. Nevertheless, also in this case, the
spectral resolution of the \xmmsp EPIC-MOS cameras is not good enough
to allow to distinguish this line from the Fe-L complex, which has an
extremely powerful emission in this temperature range.  The importance
of the Magnesium lines is rapidly decreasing. Silicon, instead is
present in a more isolated region of the spectrum (close to 2 keV),
therefore this guarantees a good measurement of its abundance for
systems with temperature up to approximately 8 keV. Another important
property of the energy region around 2 keV is that different
temperature spectra with the same normalization have there the same
flux. This means that the Silicon measurement is not significantly
influenced by the presence of a multi-temperature plasma or, in other
words, by the uncertainties related to the continuum determination.

It is important to  stress that the possibility of accurately
determining the emission lines is subject not only to the mutual
interaction of different elements, but also to the capability of
measuring with accuracy the Iron lines (which can strongly influence
Oxygen and Magnesium) and the continuum.
 
%---------------------%---------------------%---------------------% 
\section{DISCUSSION} 
\label{sec:discussion}

%---------------------%---------------------%---------------------% 
\subsection{Differences between the methods used in the spectral analysis} 
 
Considering the results presented in Sect.~\ref{sec:results} (see also
Fig.~\ref{fig:prof1}), we can conclude that the two strategies
followed in the X-ray analysis give almost identical estimates in the
case of temperature and Iron. The only relevant difference is that in
Method 2 we freeze the temperature to compute the Iron.  The effect is
to eliminate one free parameter, reducing the degeneracy, which,
however, is extremely weak in our case, because of the large number of
counts per each radial bin.  The most evident differences between the
two procedures are for the abundance measurements of the $\alpha$--metals
(Oxygen, Silicon and Magnesium): in all cases we obtain a better
result when we fix the continuum and the Iron contribution and we fit
the spectrum in a narrow band centered on the corresponding set of
lines.  We made several tests to individuate the optimal energy band
to adopt.  At the assumed redshift, i.e. $z=0.06$, the best choices
are [0.4-1.5] keV for Oxygen and [1.2-3.2] keV for both Magnesium and
Silicon.  Notice that the bands need to be large enough to include
part of the continuum to better estimate the line emission and to
evaluate the effects of each single element on the continuum.  In
particular this shrewdness is very important for Oxygen, for which the
continuum contribution is relevant on a wide energy range (see the
green line in the first panel of Fig.~\ref{fig:spec}).  Using Method 2
instead of Method 1 allows to avoid an overestimated result for
Oxygen. Moreover it permits to detect Magnesium which would not be
recovered without freezing the Iron. Finally it increases the
spectroscopic determination of Silicon avoiding a possible
underestimate.
 
%---------------------%---------------------%---------------------% 
 
\subsection{Iron overestimate for  systems at 2-3 keV} 
\label{sec:iron} 

\begin{figure*}
\begin{center}
\includegraphics[width=1.\textwidth]{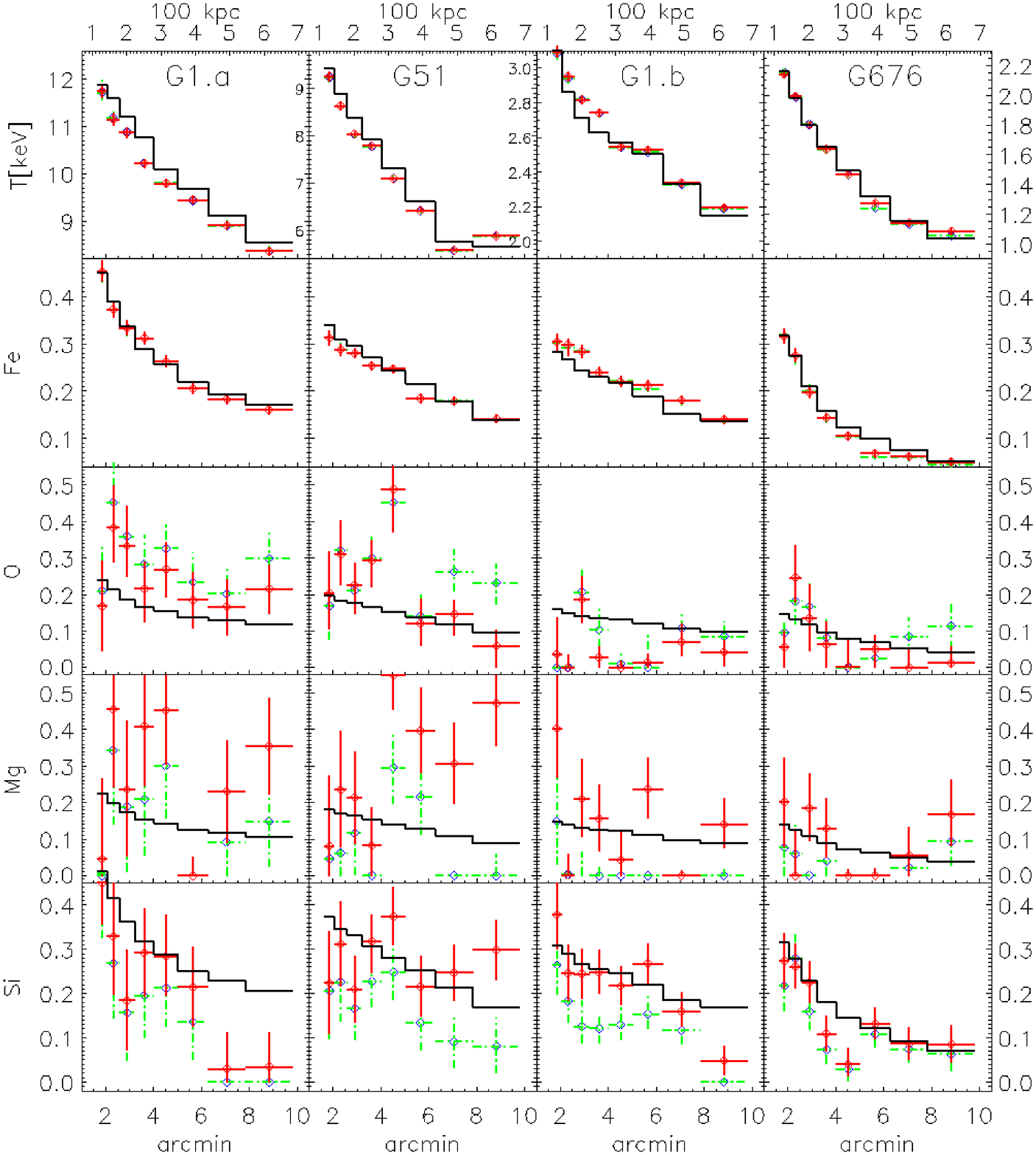}
\caption{Comparison of the spectroscopic and projected input radial profiles. 
 From top to bottom we show the
 Temperature, Iron, Oxygen, Magnesium, Silicon profiles, respectively.
Each column correspond to a different mock cluster  (from left right to
right G1.a, G51, G1.b and G676 clusters). Solid black line represents
the spectroscopic-like temperature and the emission-weighted metal
abundance. Green dashed and solid red points with relative error bars refer to
the spectroscopic values derived by using Methods 1 and 2,
respectively (see text).}
\label{fig:prof1} 
\end{center}
\end{figure*}

\begin{figure}
\begin{center}
\includegraphics[width=0.5\textwidth]{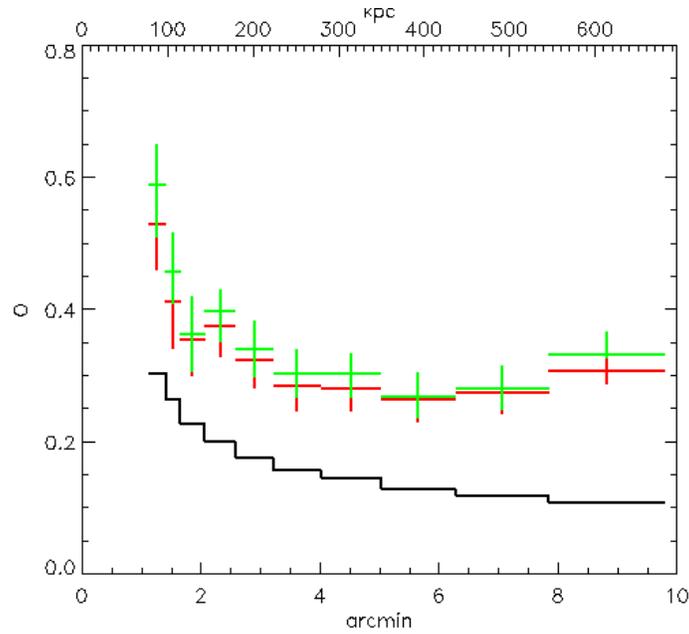}
\caption{The radial profile of Oxygen in  G1.a observed with an exposure time of 1 Ms. The meaning 
of the lines is as in Fig.~\ref{fig:prof1}.}
\label{fig:oxi}
\end{center} 
\end{figure}

\begin{figure}
\begin{center}  
\includegraphics[width=0.5\textwidth]{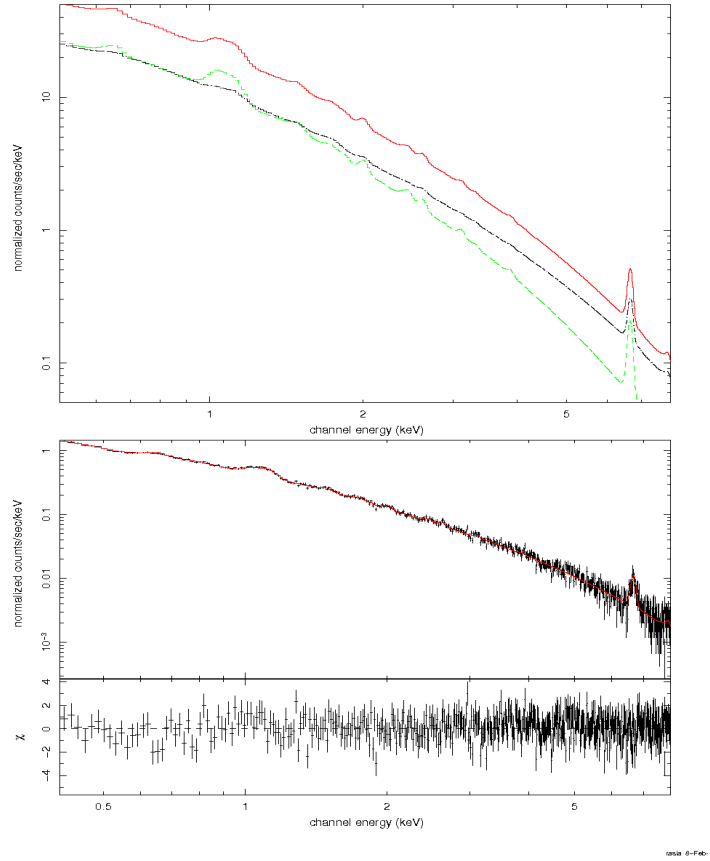}
\caption{Top panel: Spectra of a plasma at $T_1$=1 keV (green line), $T_2=$4
keV (black line) and their combination (red line). Bottom panel: a
fake spectrum with 450.000 counts obtained by combining a plasma with
$T_1=2$ keV and $Z_1=0.2$ and a plasma with $T_2=3$ keV and
$Z_2=0.1$. The red line shows the resulting best-fitting model (see
text for more details). The lowest box presents the corresponding
residuals.}
\label{fig:iron}
\end{center}   
\end{figure}

The Iron content of the ICM is perfectly recovered for hot systems,
G1.a and G51, because their Fe-K lines are extremely well determined
(see Fig.~\ref{fig:prof1} and the last 3 panels of
Fig.~\ref{fig:spec}).  Iron is also accurately estimated for cold
systems (only exception are some bins of G676 in Fig.~\ref{fig:prof1} --
see Sect.~\ref{sec:ironbias}), since their Fe-L
lines are extremely strong (see the first two panels of
Fig.~\ref{fig:spec}).  The only system for which we obtain a
systematic overestimate in all radial bins is G1.b, the 2-3 keV
object.  Notice that this discrepancy is significant, as evident from
the error bars in the corresponding panel of Fig.~\ref{fig:prof1}.
The problem is due to the fact that the G1.b temperature is very close
to the one where there is a transition between the relative
importance of the lines (Fe-L or Fe-K) used in the determination of
the global Iron content.  Moreover, its temperature structure is quite
complex: consequently inside the same radial bin we are averaging
different temperatures which sometimes are characterized by a large
Fe-K contribution, while other times show a strong Fe-L. Summing over
spectra with both high Fe-L and high Fe-K has the net effect of
increasing the final amount of Iron.  This effect can be explained by
looking at Fig.~\ref{fig:iron}.  In the first panel we combine
together a plasma at 1 keV (green line) with a plasma at 4 keV (dark
line): the resulting spectrum is shown by the red line. It is evident
that the first plasma contributes to bump the Fe-L lines, while the
second one has the role of bursting the Fe-K lines in the resulting
global spectrum, which is then characterized by an Iron content which
is larger than what expected considering its temperature.

In order to quantify this effect we perform a test with a likely
composition of the gas (see the lower panel of the same figure).  The
input data is the spectrum resulting from the combination of two
plasmas: the first one with temperature $T_1 =2 $ keV, metallicity
$Z_1 = $ 0.2 and normalization $K_1 = 1$; the second one with $T_2=3$
keV, $Z_2= 0.1$ and $K_2 = 1$.  We produce few spectra with different
number counts and fit them with a C statistics. The results are
reported in Table~\ref{tab:iron}. They show a perfect consistency for
the temperature, but a significant overestimate of the Iron spectral
value with respect to the emission-weighted one. The discrepancy is
highly significant and depends on the number counts: for 4.500 counts
the disagreement is of almost 40\%. Increasing the statistic number
the difference is alleviated but not solved.  Even observing the
cluster for $2 \times 10^{6}$ sec and thus having $9\times 10^{5}$
counts \citep[which is 1-2 orders of magnitude above the typical observed
counts rate, see e.g.][]{balestra.etal.07}, the discrepancy is still
present and of order of 15\%. The spectrum corresponding to the
best-fit (red line) of the $4.5 \times 10^{5}$ counts spectrum is compared to
the input data in the same panel of Fig.\ref{fig:iron}, while in the
bottom box we show the corresponding residuals.  Notice that in the
residual behavior there are not evident features which can make us
suspicious that the fit is not good or that it can be affected by some
bias on the Iron lines.

The implications of these results are extremely important. Iron is in
fact the metal to which all the other $\alpha$--elements are
referring, thus overestimating its content can lead to wrong
conclusions on the past stellar population.  

This effect may also explain, at least partially, the recent results
obtained by \cite{baumgartner.etal.05}. After stacking in temperature
bins 273 clusters observed with \asca, they studied the global metal
abundances as a function of the mean temperature. They found that in
all temperature bins (from 0.5 keV to 12 keV) the emission-weighted
Iron abundance within the cluster region selected to contain as much
flux as possible is around 0.2-0.25 (in solar units), with the only
exception of the two bins between 2 and 4 keV, where a global Iron
value of 0.43--0.47 $Z_{\odot}$ is measured.

 %---------------------%---------------------%---------------------% 
 \subsection{Fe bias for cold systems}
 \label{sec:ironbias}
 
Another bias affecting the estimate of the Iron content is present in
some bins of the cold system G676 (see Fig.~\ref{fig:prof1}). The
systematic underestimate is know in literature as "Iron bias"
\citep{buote00a,buote00b}. However, this bias 
gives at maximum an underestimate of order of 20\% and it is present
only in G676, but it does not influence the other two cold
systems. The poor evidence of this bias in our sample can be explained
as follow.

The Iron bias introduced by \cite{buote00a} is caused by the fact of
pretending to fit with a single temperature model a plasma which is,
instead, characterized either by a combination of different
temperatures (multi-temperature plasma) or by a strong temperature
gradient. Observationally, the bias is seen to affect the central
cooling core regions where typically there is a positive
temperature gradient ($dT/dr >0$) as well as both negative metallicity
gradient ($dZ/dr <0$) and negative emission gradient ($dEM/dr<0$). The
combination of these factors leads to enlarge the bump of Fe-L shell
(the colder plasma with higher metallicity and emission excites the
lines on the soft part of Fe-L shell). In our work, we are not in this
regime. On the contrary, both temperature and metallicity gradient are
shallow inside the spatial bins we considered. Moreover the
temperature gradient in the our external regions is going in the
opposite direction ($dT/dr < 0$) with respect its behavior in the
cooling regions.
 
Another reason for which we do not detect a strong Fe bias in our
system is because it decreases if the spectral fitting is made in an
energy band sufficiently large to determine the continuum or, in other
words, the temperature profile
\citep{buote00b} with accuracy. \cite{buote.etal.03} suggested to
extend the minimum of the fitting energy band down below 0.6 keV in
order to limit the effect of this bias.  With our analysis we respect
this demand since the minimum energy considered is 0.4 keV. In
addition, a robust determination of the continuum is guaranteed by the
high signal to noise that we have up to 4 keV.  The fact that a good
Iron measurement depends on a correct temperature measurement is clear
comparing in Fig.~\ref{fig:prof1} the G676 temperature and Iron panels.

%---------------------%---------------------%---------------------% 
\subsection{Oxygen and Magnesium for hot clusters} 
 
In Sect.~\ref{sec:results} we showed that the abundance estimates for
both Oxygen and Magnesium do not show any problem of systematics for
systems colder than 2 keV.  On the other hand, for the hot clusters,
like G1.a and G51, we found a clear bias producing a significant
overestimate.  Notice that these elements are the best
$\alpha$--element indicators, thus they are a fundamental tool for
studying the stellar evolution and formation history through the ratio
between SNe Ia and SNe II.  It is therefore necessary to understand in
detail the origin of the discrepancy. At this aim we investigate a
number of possible reasons which depend on the ICM physics implemented
in the original hydrodynamic simulation, on the complexity of the
plasma temperature and on the features of the X-ray spectra.
 
\begin{figure}
\begin{center}  
\includegraphics[width=0.5\textwidth]{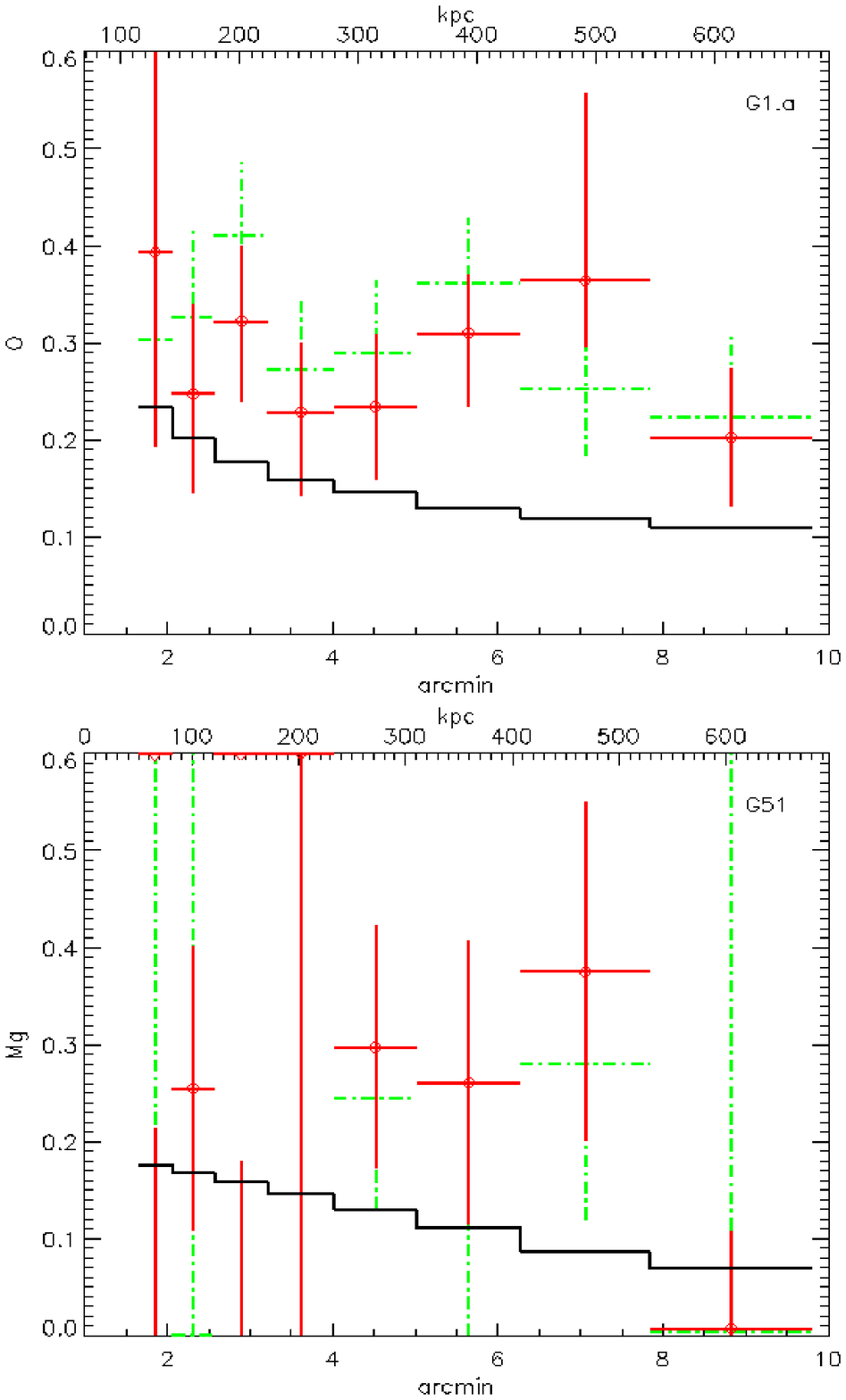}
\caption{
The radial profiles for Oxygen in G1.a (top panel) and Magnesium in
G51 (bottom panel), obtained considering only particles with
temperature larger than 5 keV.  Symbols and lines are as in
Fig.~\ref{fig:prof1}.}
\label{fig:oxi2}
\end{center}   
\end{figure}

\begin{itemize} 
 
\item {\bf Presence of small undetected cold blobs}.  
Clusters G1.a and G51 show in their photon images
(Fig.~\ref{fig:clusters}) a large number of compact gas clouds spread
in the ICM which are usually associated to cold blobs in the
temperature maps (Fig.~\ref{fig:tsl}).  Most of them have been
detected by the wavelet decomposition algorithm and then excluded from
our analysis, but some of them are still present, in
particular in the external regions.  These structures, which have a
very low temperature, could in principle be responsible for an overestimate for
Oxygen or Magnesium, since their importance is greater for cold
systems.  To investigate this problem in more detail we generate new
event files for G1.a and G51, but using only the particles having a
temperature larger of 5 keV: in this way we are sure to avoid any
contamination from cold (diffuse and clumped) gas.  To these event
files we apply the whole procedure describe above.  As expected, using
the wavelength algorithm, we do not find any compact cool source and we just
exclude the second brightest blob, located on the East side of G51
centre.  The comparison between the spectroscopic results and the
simulation profiles of Oxygen for G1.a and Magnesium for G51
(re-computed excluding the particles with temperature smaller than 5
keV) are shown in Fig.~\ref{fig:oxi2}: it is clear that the
discrepancies are not alleviated.

\item {\bf Dynamical state of the cluster}. 
Since the bias is evident only for the two largest clusters which
clearly have a complex morphological structure, while it does not
affect the coolest systems which are more symmetric and regular, its
relevance could be enhanced by the presence of thermal inhomogeinities
in the ICM.  However, we notice that G1.b does not show this
phenomenon, even if it represents a perturbed object.  To further
investigate this possibility, we generate another series of events
files fixing the temperature of all the particles at a given constant
value (8 keV) but letting all the metals equal to their original
values. In this way we create an object with a completely flat
temperature structure.  We repeat the analysis on these new images,
finding again the same discrepancy for the Oxygen and Magnesium
abundances.
 
\item {\bf Fe-L over Fe-K ratio}. 
The spectra we have extracted from the images are the sum of
contributions coming from the plasma present along the line of sight
and having different temperatures and metallicities.  The resulting
spectra, therefore, could have a Fe-L over Fe-K ratio different from
the typical value reported by the \vmekal model in correspondence of
the measured temperature. The Iron content of hot systems is estimated
basically using the Fe-K systems which are strongly more powerful than
the Fe-L lines at those temperatures. The Fe-L group of the spectra can
be, thus, weaker than what we expect from a \vmekal model where we
estimate the Iron content by fitting the Fe-K lines.  In this
situation one way to ``fill the gap'' between the true Fe-L of the
spectra and the expected one of the model is to increase the emission
power by other elements present in the same energy band.  We performed
a detailed comparison of the spectral Fe-L and Fe-K measurements and
the corresponding model expectations. We conclude that the Magnesium
estimate can be affected for this reason. In fact for G51, which is a
system showing a large disagreement for Mg lines, the Iron
determination from the Fe-L is always higher than what we derive from
the Fe-K lines. On the other hand, this behavior is not present in
G1.a and thus cannot explain the disagreement found for Oxygen, which
is also displaced by Fe-L at temperatures larger than 8 keV (see
Fig.~\ref{fig:spec}).
 
\item{\bf Continuum determination}.  

Oxygen is a weak line at temperatures larger than 5 keV, thus a wrong
determination of the continuum can strongly influence the measurement
of this element. In particular an underestimate of the continuum can
originate an overestimate of the Oxygen abundance.  To have a rough
estimate of this effect we simulate a spectrum as sum of a 6 keV
plasma with Oxygen equal to 0.1 (in solar units) and a 8 keV plasma
with Oxygen fixed to 0.2. The two plasmas have been weighted in the
same way by assigning a unity normalization to both.  The fake
spectrum has been convolved with the \xmmsp response and simulated
assuming an exposure time of 1 Msec. This large integration time
allows us to have few millions of counts per spectra.  We performed a
first fit using a $\chi^2$ statistics and a \vmekal model where we let
free temperature, Oxygen and normalization: the resulting temperature
agrees with the spectroscopic-like estimate.  This temperature has
been then fixed in the following tests, where we also fixed the
normalization, but to different values, starting the input theoretical
value of 2 and then decreasing it slightly every time.  In this way we
are able to study the effect that only a wrong determination of the
continuum has on the spectroscopic determination of Oxygen.  The
expected emission-weighted abundance for Oxygen is 0.149.  This value
is perfectly recovered when we force the normalization to be the
theoretical one. A result compatible at 1$\sigma$ level is still
obtained using a lower normalization, up to 1.995.  Decreasing further
the normalization by 0.5 or 0.75 per cent, the spectroscopic
determination of Oxygen becomes much larger than the expected one.
This simple test shows us how Oxygen is very dependent on the goodness
of the continuum determination and suggests that the measurements done
through \xmmsp in high-temperature systems can be significantly biased.
\end{itemize} 
 
%-------------- 
 
%%%%%%%%%%%%%%%%%%%-----------------------------%%%%%%%%%%%%%%%%%%% 

\section{Conclusions}   
\label{sec:conclu} 

The Intracluster Medium represents a perfect laboratory  to study
the physics of all the cluster subcomponents, like its galaxies and
the stars in galaxies.  In particular the ICM metallicity is a rich
source of information to test the stellar evolution models.  In this
work we deeply investigated all possible systematic biases present in
the metallicity measurements from X-ray spectra having the same
spectral resolution of \xmm. At this aim we improved the X-ray MAp
Simulator in order to correctly consider the metal information which
is provided by our simulations. The six clusters here analyzed cover a
wide range of temperature and have different dynamical state.  This
allows us to determine how the biases we analyzed are effectively
influenced by the complexity of the thermal structure and by the value
of the temperature itself.

Performing the X-ray analysis we tested two standard procedures
adopted by X-ray observers: first, we fitted the spectra in a large
band leaving free at the same time all the parameters of interest
($T$, Fe, O, Mg, Si); second, we recover the temperature and the Iron
content from a fit in a large band and then we measure the
$\alpha$--elements, O, Mg, and Si, in a narrower band centered on the
lines.  Our main results are shown in Fig.~\ref{fig:prof1}
and can be summarized as follows:
\begin{itemize} 
\item[-] Using narrow bands to measure all the elements beside Iron is 
in general better for all clusters, independently of their
temperature. This procedure prevents to overestimate Oxygen and to
underestimate Silicon. Furthermore it increases the probability to
detect the Magnesium content.
\item[-] Iron is perfectly recovered as soon as either  Fe-K or Fe-L are 
the dominating lines: this occurs when the plasma has a temperature
larger than 3 keV or smaller than 2 keV, respectively.  In our sample
we have an object showing a complex map temperature
(Fig.~\ref{fig:tsl}) with values ranging between 2-3 keV (see its
temperature profile in Fig.~\ref{fig:prof1}). Simulating a simple
model which describes this system in \xspec we found that the
spectroscopic Iron is overestimated by up to 15-40 per cent depending
on the number statistic: decreasing the source counts leads to an
increase of the disagreement. This result has strong implications. In
fact one of the main goals of measuring the ICM metallicity is the
derivation of the [$\alpha$/Fe] and [Si/Fe] ratios (see the
Introduction).  Therefore, the fact that we detect a clear systematic
effect in the Iron measurements can affect the results that we can
derive from them about the stellar formation history.

\item[-] Oxygen is well measured for clusters with temperature lower 
than 3 keV, while we find an evidence of overestimate for the hottest
system in our sample, which has a temperature profile ranging from 8
to 12 keV (see Fig.~\ref{fig:prof1}). We check carefully some of the
possible reasons that can originate this effect. The explanation came
out to be the problematic balancing between the determination of the
continuum and the weakness of this line at temperature higher than 8
keV. A discrepancy smaller than 1 per cent on the normalization is
sufficient to explain the Oxygen overestimate.
 
\item[-] The measurement of  Magnesium is difficult  at all temperatures:
 for cold objects (with $T<2$ keV) its line is, in fact, lying in the
same spectral range of strong Fe-L lines and thus its detection is
hidden, while at high temperatures it is a very weak line.  We found a
systematic overestimate of Magnesium in hot clusters: this effect can be
related to the fact that the [Fe-L/Fe-K] ratio is higher in the
spectra than in the model used to fit them. Thus to fill the Fe-L
group lines there is the necessity to pump the emission of other
elements present in the same energy band, that in our case, is
Magnesium.
 
\item[-] Finally, from the X-ray analysis of the spectra we can measure with 
good precision Silicon for all the objects in our sample
independently of their dynamical state.  This is due to the fact that
the value of the spectra around 2 keV is only slightly dependent on
the temperature.
\end{itemize} 
 
The method used in this paper to investigate the metallicity
distribution takes an enormous advantage by the use of high-resolution
simulations where the hydrodynamic physics is treated including many
different physical processes which are relevant for the thermal and
chemical evolution of the ICM.  Other aspects which can play a
significant role, like AGNs, cosmic rays and magnetic field, are not
included in these simulations but could be added in the next
future. At the present the state of our simulations does not
fully reproduce the cluster core, where more of the X-ray information
comes from, but it is a proper description of the temperature gradient
outside 0.1 $R_{200}$ \citep{pratt.etal.07}.  Nevertheless, it is
worth stressing that our results and the corresponding explanations
are not dependent on the kind of simulations used.  In fact, when we
found some disagreement between the input profiles from our simulation
and the X-ray spectroscopic measurements, we investigated the problem
by analyzing the spectral properties of fake plasma spectra generated
using
\xspec. For this reason our results can be considered valid for observations
of spectra generated by a plasma with a multi-component temperature.
The only restrictions of our results are due to the fact that we
focused our effort on a fixed spectral energy resolution, like the
\xmmsp one.   The powerful of the analysis of the ICM metal content can
radically change in the next future with an X-ray satellite having a
micro-calorimeter on board.  EDGE (Explorer of Diffuse Emission and
Gamma-ray burst Explosions)
\footnote{http://projects.iasf-roma.inaf.it/edge/} and
Constellation-X\footnote{http://constellation.gsfc.nasa.gov/index.html}
are two projects that have been proposed with the intent of
incorporating such technology, that will allow to solve the blending
of Fe-L and Oxygen lines.
  
%%%%%%%%%%%%%%%%%%%-----------------------------%%%%%%%%%%%%%%%%%%% 
\section*{ACKNOWLEDGMENTS}  
  
We thank the referee for useful comments. 

The simulations have been performed with CPU time allocated at the
``Centro Interuniversitario del Nord-Est per il Calcolo Elettronico''
(CINECA, Bologna) thanks to grants from INAF and from the University
of Trieste. Support for this work was provided by the INFN grant PD-51
and by NASA through Chandra Postdoctoral Fellowship grant number
PF6-70042 awared by the Chandra X-ray Center, which is operated by the
Smithsonian Astrophysical Observatory for NASA under contract
NAS8-03060. We acknowledge the financial support from contract
ASI-INAF I/023/05/0. ER thanks for the hospitality the MPA in
Garching, where part of the analysis was done.  PM acknowledge support
from NASA grants GO4-5155X and GO5-6124X.  We are extremely grateful
to Jean-Luc Sauvageot for sharing with us a previous version of a code
which produced \xmmsp images; to Mauro Roncarelli, Eugenio Ursino and
Massimiliano Galeazzi for discussing on the new distribution
technique; to Alexey Vikhlinin for providing information on the
wavelet detection algorithm; to Kyoko Matsushita and Fabio Gastaldello
for kindly discussions on X-ray metallicity analysis.

\newpage

%-------------------- 
%-------------------- 
%-------------------- 

%-------------------- 
%-------------------- 
%---------------

%--------------

%---------------
%---------------

%=-=-=-=-=-=-TABLE=-=-=-=-=-=-

\begin{table} 
\caption{The main physical properties of the simulated clusters analyzed in this paper:
R$_{200}$ and $R_{500}$ in units of Mpc$/h$ (the radius of the spheres
with a density 200 and 500, respectively, times the critical density
of the Universe); the masses inside these radii, $M_{200}$ and
$M_{500}$, in units of $10^{14} M_{\odot}/h$, and the mass-weighted
temperatures in keV, $T_{\rm MW,200}$.}
 
\begin{tabular}{l c c c c c}  
 & & \\ 
% TMW calcolata con smac mettendodentro i vari raggi tipici della tabella 
 & $R_{\rm 200}$ & $R_{\rm 500}$ & $M_{\rm 200}$ & $M_{\rm 500}$& $T_{\rm MW,200}$ \\ 
\hline \\  
G1.a  & 1.771 & 1.202 & 12.92 & 10.09 & 7.37  \\ 
G51   & 1.675 & 1.120 & 10.92 &  8.16 & 6.44  \\ 
G1.b  & 1.151 & 0.740 &  3.55 &  2.36 & 2.63  \\ 
G676  & 0.727 & 0.490 &  0.89 &  0.68 & 1.42  \\ 
G914  & 0.718 & 0.485 &  0.86 &  0.66 & 1.39  \\ 
G1542 & 0.725 & 0.484 &  0.88 &  0.66 & 1.32  \\ 
 
%&&&&\\ 
\end{tabular}  
\label{tab:clusters} 
\end{table}  

\begin{table} 
\caption{Values of Oxygen abundances (with their 1$\sigma$ error bars), 
derived by fitting a spectra, which is the combination of two plasmas
at 6 and 8 keV (more details in the text).  In the $\chi^2$ fitting
procedure we fix the temperature at the spectroscopic-like value and
then we fix also the normalization, considering the its theoretical
value (2) and smaller values (1.995, 1.990, 1.985).  The percentages
(within brackets) represent the difference between the values
recovered from the spectral analysis and the expected
emission-weighted one: O=0.149.}
\begin{tabular}{l c c c c}   
 & & \\ & $N$= 2.000 & 1.995 & 1.990 & 1.985 \\
\hline \\   
O$_{\rm min}$ & 0.126  & 0.148       & 0.175 (15\%)& 0.192(29\%) \\
O             & 0.141  & 0.162 (9\%) & 0.184 (23\%)& 0.198(38\%) \\
O$_{\rm max}$ & 0.155  & 0.176 (18\%)& 0.198 (33\%)& 0.220(48\%) \\
 &&&\\ 
 $\chi^2_{Red}$ (70 d.o.f.) & 1.08  & 1.08  & 1.12 & 1.20 \\  
\end{tabular}   
\label{tab:oxi}  
\end{table}

\begin{table} 
\caption{Values of Temperature and Iron abundances (with their 1$\sigma$ error
bars), derived by fitting a spectra of a T=2 keV and Z=0.2 plasma
combined with another plasma of T=3 keV and Z=0.1.  The percentages
(within brackets) represent the difference between the values
recovered from the spectral analysis and the expected
emission-weighted one: Fe=0.144. The expected spectroscopic-like
value of the temperature is equal to 2.45.}
\begin{tabular}{l c c c c c c }   
 & & \\ 
$\# counts$    &4.5$10^3$ &9$10^3$ &4.5$10^4$& 9$10^4$ & 4.5$10^5$ & 9$10^5$\\
\hline \\   
T$_{\rm min}$  &2.361     &2.380   &2.484    &  2.426  & 2.459     & 2.448 \\
T              &2.439     &2.434   &2.510    &  2.444  & 2.470     & 2.454 \\
T$_{\rm max}$  &2.520     &2.490   &2.536    &  2.462  & 2.476     & 2.460 \\
 &&&\\ 
Fe$_{\rm min}$ &0.158(10\%)&0.158(10\%)&0.172(19\%)&0.174(21\%)&0.170(18\%)&0.162(12\%)\\
Fe             &0.199(38\%)&0.186(29\%)&0.185(28\%)&0.183(27\%)&0.174(21\%)&0.165(16\%)\\
Fe$_{\rm max}$ &0.243(69\%)&0.215(49\%)&0.198(37\%)&0.192(33\%)&0.178(24\%)&0.168(17\%)\\

\end{tabular}   
\label{tab:iron}  
\end{table}


\newpage

\begin{thebibliography}{74}
\expandafter\ifx\csname natexlab\endcsname\relax\def\natexlab#1{#1}\fi

\bibitem[{{Anders} \& {Grevesse}(1989)}]{anders&grevesse}
{Anders}, E. \& {Grevesse}, N. 1989, \gca, 53, 197

\bibitem[{{Arnaud}(1996)}]{arnaud96}
{Arnaud}, K.~A. 1996, in ASP Conf. Ser. 101: Astronomical Data Analysis
  Software and Systems V, 17--+

\bibitem[{{Balestra} {et~al.}(2007){Balestra}, {Tozzi}, {Ettori}, {Rosati},
  {Borgani}, {Mainieri}, {Norman}, \& {Viola}}]{balestra.etal.07}
{Balestra}, I., {Tozzi}, P., {Ettori}, S., {Rosati}, P., {Borgani}, S.,
  {Mainieri}, V., {Norman}, C., \& {Viola}, M. 2007, \aap, 462, 429

\bibitem[{{Baumgartner} {et~al.}(2005){Baumgartner}, {Loewenstein}, {Horner},
  \& {Mushotzky}}]{baumgartner.etal.05}
{Baumgartner}, W.~H., {Loewenstein}, M., {Horner}, D.~J., \& {Mushotzky}, R.~F.
  2005, \apj, 620, 680

\bibitem[{{Bertone} {et~al.}(2007){Bertone}, {De Lucia}, \&
  {Thomas}}]{bertone.etal.07}
{Bertone}, S., {De Lucia}, G., \& {Thomas}, P.~A. 2007, preprint,
  astro-ph/0701407

\bibitem[{{B{\"o}hringer} {et~al.}(2004){B{\"o}hringer}, {Matsushita},
  {Churazov}, {Finoguenov}, \& {Ikebe}}]{Boehringer.etal.04}
{B{\"o}hringer}, H., {Matsushita}, K., {Churazov}, E., {Finoguenov}, A., \&
  {Ikebe}, Y. 2004, \aap, 416, L21

\bibitem[{{B{\"o}hringer} {et~al.}(2002){B{\"o}hringer}, {Matsushita},
  {Churazov}, {Ikebe}, \& {Chen}}]{boehringer.etal.02}
{B{\"o}hringer}, H., {Matsushita}, K., {Churazov}, E., {Ikebe}, Y., \& {Chen},
  Y. 2002, \aap, 382, 804

\bibitem[{{B{\"o}hringer} {et~al.}(2005){B{\"o}hringer}, {Matsushita},
  {Finoguenov}, {Xue}, \& {Churazov}}]{boehringer.etal.05}
{B{\"o}hringer}, H., {Matsushita}, K., {Finoguenov}, A., {Xue}, Y., \&
  {Churazov}, E. 2005, Advances in Space Research, 36, 677

\bibitem[{{Bourdin} {et~al.}(2004){Bourdin}, {Sauvageot}, {Slezak}, {Bijaoui},
  \& {Teyssier}}]{bourdin.etal.04}
{Bourdin}, H., {Sauvageot}, J.-L., {Slezak}, E., {Bijaoui}, A., \& {Teyssier},
  R. 2004, \aap, 414, 429

\bibitem[{{Buote}(2000a)}]{buote00a}
{Buote}, D.~A. 2000a, \mnras, 311, 176

\bibitem[{{Buote}(2000b)}]{buote00b}
{Buote}, D.~A. 2000b, \apj, 539, 172

\bibitem[{{Buote} {et~al.}(2003){Buote}, {Lewis}, {Brighenti}, \&
  {Mathews}}]{buote.etal.03}
{Buote}, D.~A., {Lewis}, A.~D., {Brighenti}, F., \& {Mathews}, W.~G. 2003,
  \apj, 595, 151

\bibitem[{{Cora}(2006)}]{cora06}
{Cora}, S.~A. 2006, \mnras, 368, 1540

\bibitem[{{De Grandi} {et~al.}(2004){De Grandi}, {Ettori}, {Longhetti}, \&
  {Molendi}}]{degrandi.etal.04}
{De Grandi}, S., {Ettori}, S., {Longhetti}, M., \& {Molendi}, S. 2004, \aap,
  419, 7

\bibitem[{{De Grandi} \& {Molendi}(2001)}]{degrandi&molendi01}
{De Grandi}, S. \& {Molendi}, S. 2001, \apj, 551, 153

\bibitem[{{De Lucia} {et~al.}(2004){De Lucia}, {Kauffmann}, \&
  {White}}]{delucia.etal.04}
{De Lucia}, G., {Kauffmann}, G., \& {White}, S.~D.~M. 2004, \mnras, 349, 1101

\bibitem[{{de Plaa} {et~al.}(2007){de Plaa}, {Werner}, {Bleeker}, {Vink},
  {Kaastra}, \& {M{\'e}ndez}}]{deplaa.etal.07}
{de Plaa}, J., {Werner}, N., {Bleeker}, J.~A.~M., {Vink}, J., {Kaastra}, J.~S.,
  \& {M{\'e}ndez}, M. 2007, \aap, 465, 345

\bibitem[{{de Plaa} {et~al.}(2006){de Plaa}, {Werner}, {Bykov}, {Kaastra},
  {M{\'e}ndez}, {Vink}, {Bleeker}, {Bonamente}, \& {Peterson}}]{deplaa.etal.06}
{de Plaa}, J., {Werner}, N., {Bykov}, A.~M., {Kaastra}, J.~S., {M{\'e}ndez},
  M., {Vink}, J., {Bleeker}, J.~A.~M., {Bonamente}, M., \& {Peterson}, J.~R.
  2006, \aap, 452, 397

\bibitem[{{Dolag} {et~al.}(2005){Dolag}, {Vazza}, {Brunetti}, \&
  {Tormen}}]{dolag.etal.05a}
{Dolag}, K., {Vazza}, F., {Brunetti}, G., \& {Tormen}, G. 2005, \mnras, 994

\bibitem[{{Domainko} {et~al.}(2006){Domainko}, {Mair}, {Kapferer}, {van
  Kampen}, {Kronberger}, {Schindler}, {Kimeswenger}, {Ruffert}, \&
  {Mangete}}]{domainko.etal.06}
{Domainko}, W., {Mair}, M., {Kapferer}, W., {van Kampen}, E., {Kronberger}, T.,
  {Schindler}, S., {Kimeswenger}, S., {Ruffert}, M., \& {Mangete}, O.~E. 2006,
  \aap, 452, 795

\bibitem[{{Dupke} \& {White}(2003)}]{dupke&white03}
{Dupke}, R. \& {White}, III, R.~E. 2003, \apjl, 583, L13

\bibitem[{{Ettori} {et~al.}(2002){Ettori}, {Fabian}, {Allen}, \&
  {Johnstone}}]{ettori.etal.02b}
{Ettori}, S., {Fabian}, A.~C., {Allen}, S.~W., \& {Johnstone}, R.~M. 2002,
  \mnras, 331, 635

\bibitem[{{Finoguenov} {et~al.}(2001){Finoguenov}, {Arnaud}, \&
  {David}}]{finoguenov.etal.01a}
{Finoguenov}, A., {Arnaud}, M., \& {David}, L.~P. 2001, \apj, 555, 191

\bibitem[{{Finoguenov} {et~al.}(2000){Finoguenov}, {David}, \&
  {Ponman}}]{finoguenov.etal.00}
{Finoguenov}, A., {David}, L.~P., \& {Ponman}, T.~J. 2000, \apj, 544, 188

\bibitem[{{Finoguenov} \& {Jones}(2000)}]{finoguenov&jones00}
{Finoguenov}, A. \& {Jones}, C. 2000, \apj, 539, 603

\bibitem[{{Finoguenov} {et~al.}(1999){Finoguenov}, {Jones}, {Forman}, \&
  {David}}]{finoguenov.etal.99}
{Finoguenov}, A., {Jones}, C., {Forman}, W., \& {David}, L. 1999, \apj, 514,
  844

\bibitem[{{Finoguenov} {et~al.}(2002){Finoguenov}, {Matsushita},
  {B{\"o}hringer}, {Ikebe}, \& {Arnaud}}]{Finoguenov.etal.02}
{Finoguenov}, A., {Matsushita}, K., {B{\"o}hringer}, H., {Ikebe}, Y., \&
  {Arnaud}, M. 2002, \aap, 381, 21

\bibitem[{{Finoguenov} \& {Ponman}(1999)}]{finoguenov&ponman99}
{Finoguenov}, A. \& {Ponman}, T.~J. 1999, \mnras, 305, 325

\bibitem[{{Fukazawa} {et~al.}(1998){Fukazawa}, {Makishima}, {Tamura}, {Ezawa},
  {Xu}, {Ikebe}, {Kikuchi}, \& {Ohashi}}]{fukazawa.etal.98}
{Fukazawa}, Y., {Makishima}, K., {Tamura}, T., {Ezawa}, H., {Xu}, H., {Ikebe},
  Y., {Kikuchi}, K., \& {Ohashi}, T. 1998, \pasj, 50, 187

\bibitem[{{Fukazawa} {et~al.}(2000){Fukazawa}, {Makishima}, {Tamura},
  {Nakazawa}, {Ezawa}, {Ikebe}, {Kikuchi}, \& {Ohashi}}]{fukazawa.etal.00}
{Fukazawa}, Y., {Makishima}, K., {Tamura}, T., {Nakazawa}, K., {Ezawa}, H.,
  {Ikebe}, Y., {Kikuchi}, K., \& {Ohashi}, T. 2000, \mnras, 313, 21

\bibitem[{{Fukazawa} {et~al.}(1994){Fukazawa}, {Ohashi}, {Fabian}, {Canizares},
  {Ikebe}, {Makishima}, {Mushotzky}, \& {Yamashita}}]{fukazawa.etal.94}
{Fukazawa}, Y., {Ohashi}, T., {Fabian}, A.~C., {Canizares}, C.~R., {Ikebe}, Y.,
  {Makishima}, K., {Mushotzky}, R.~F., \& {Yamashita}, K. 1994, \pasj, 46, L55

\bibitem[{{Gardini} {et~al.}(2004){Gardini}, {Rasia}, {Mazzotta}, {Tormen}, \&
  {Moscardini}}]{xmas}
{Gardini}, A., {Rasia}, E., {Mazzotta}, P., {Tormen}, G. aùnd~{De Grandi}, S.,
  \& {Moscardini}, L. 2004, \mnras, 351, 505

\bibitem[{{Gastaldello} \& {Molendi}(2002)}]{Gastaldello&Molendi02}
{Gastaldello}, F. \& {Molendi}, S. 2002, \apj, 572, 160

\bibitem[{{Haardt} \& {Madau}(1996)}]{haardt&madau}
{Haardt}, F. \& {Madau}, P. 1996, \apj, 461, 20

\bibitem[{{Humphrey} {et~al.}(2006){Humphrey}, {Buote}, {Gastaldello},
  {Zappacosta}, {Bullock}, {Brighenti}, \& {Mathews}}]{humphery&buote06}
{Humphrey}, P.~J., {Buote}, D.~A., {Gastaldello}, F., {Zappacosta}, L.,
  {Bullock}, J.~S., {Brighenti}, F., \& {Mathews}, W.~G. 2006, \apj, 646, 899

\bibitem[{{Lia} {et~al.}(2002){Lia}, {Portinari}, \& {Carraro}}]{lia.etal.02}
{Lia}, C., {Portinari}, L., \& {Carraro}, G. 2002, \mnras, 330, 821

\bibitem[{{Liedahl} {et~al.}(1995){Liedahl}, {Osterheld}, \&
  {Goldstein}}]{liedahl.etal.95}
{Liedahl}, D.~A., {Osterheld}, A.~L., \& {Goldstein}, W.~H. 1995, \apjl, 438,
  L115

\bibitem[{{Matsushita} {et~al.}(2007{\natexlab{a}}){Matsushita},
  {B{\"o}hringer}, {Takahashi}, \& {Ikebe}}]{matsushita.etal.07}
{Matsushita}, K., {B{\"o}hringer}, H., {Takahashi}, I., \& {Ikebe}, Y.
  2007{\natexlab{a}}, \aap, 462, 953

\bibitem[{{Matsushita} {et~al.}(2003){Matsushita}, {Finoguenov}, \&
  {B{\"o}hringer}}]{Matsushita.etal.03}
{Matsushita}, K., {Finoguenov}, A., \& {B{\"o}hringer}, H. 2003, \aap, 401, 443

\bibitem[{{Matsushita} {et~al.}(2007{\natexlab{b}}){Matsushita}, {Fukazawa},
  {Hughes}, {Kitaguchi}, {Makishima}, {Nakazawa}, {Ohashi}, {Ota}, {Tamura},
  {Tozuka}, {Tsuru}, {Urata}, \& {Yamasaki}}]{matsushita.etal.06}
{Matsushita}, K., {Fukazawa}, Y., {Hughes}, J.~P., {Kitaguchi}, T.,
  {Makishima}, K., {Nakazawa}, K., {Ohashi}, T., {Ota}, N., {Tamura}, T.,
  {Tozuka}, M., {Tsuru}, T.~G., {Urata}, Y., \& {Yamasaki}, N.~Y.
  2007{\natexlab{b}}, \pasj, 59, 327

\bibitem[{{Matsushita} {et~al.}(2000){Matsushita}, {Ohashi}, \&
  {Makishima}}]{matsushita.etal.00}
{Matsushita}, K., {Ohashi}, T., \& {Makishima}, K. 2000, \pasj, 52, 685

\bibitem[{{Maughan} {et~al.}(2007){Maughan}, {Jones}, {Forman}, \& {Van
  Speybroeck}}]{maughan.etal.07}
{Maughan}, B.~J., {Jones}, C., {Forman}, W., \& {Van Speybroeck}, L. 2007,
  ArXiv Astrophysics e-prints, astro-ph/0703156

\bibitem[{{Mazzotta} {et~al.}(2004){Mazzotta}, {Rasia}, {Moscardini}, \&
  {Tormen}}]{tsl}
{Mazzotta}, P., {Rasia}, E., {Moscardini}, L., \& {Tormen}, G. 2004, \mnras,
  354, 10

\bibitem[{{Mewe} {et~al.}(1985){Mewe}, {Gronenschild}, \& {van den
  Oord}}]{mewe.etal.85}
{Mewe}, R., {Gronenschild}, E.~H.~B.~M., \& {van den Oord}, G.~H.~J. 1985,
  \aaps, 62, 197

\bibitem[{{Mewe} {et~al.}(1986){Mewe}, {Lemen}, \& {van den
  Oord}}]{mewe.etal.86}
{Mewe}, R., {Lemen}, J.~R., \& {van den Oord}, G.~H.~J. 1986, \aaps, 65, 511

\bibitem[{{Molendi} \& {Gastaldello}(2001)}]{molendi&gastaldello01}
{Molendi}, S. \& {Gastaldello}, F. 2001, \aap, 375, L14

\bibitem[{{Moll} {et~al.}(2007){Moll}, {Schindler}, {Domainko}, {Kapferer},
  {Mair}, {van Kampen}, {Kronberger}, {Kimeswenger}, \&
  {Ruffert}}]{moll.etal.07}
{Moll}, R., {Schindler}, S., {Domainko}, W., {Kapferer}, W., {Mair}, M., {van
  Kampen}, E., {Kronberger}, T., {Kimeswenger}, S., \& {Ruffert}, M. 2007,
  \aap, 463, 513

\bibitem[{{Monaghan} \& {Lattanzio}(1985)}]{monaghan&lattanzio}
{Monaghan}, J.~J. \& {Lattanzio}, J.~C. 1985, \aap, 149, 135

\bibitem[{{Morrison} \& {McCammon}(1983)}]{morrison&mccammon}
{Morrison}, R. \& {McCammon}, D. 1983, \apj, 270, 119

\bibitem[{{Mushotzky} {et~al.}(1996){Mushotzky}, {Loewenstein}, {Arnaud},
  {Tamura}, {Fukazawa}, {Matsushita}, {Kikuchi}, \&
  {Hatsukade}}]{mushotzky.etal.96}
{Mushotzky}, R., {Loewenstein}, M., {Arnaud}, K.~A., {Tamura}, T., {Fukazawa},
  Y., {Matsushita}, K., {Kikuchi}, K., \& {Hatsukade}, I. 1996, \apj, 466, 686

\bibitem[{{Nagashima} {et~al.}(2005){Nagashima}, {Lacey}, {Baugh}, {Frenk}, \&
  {Cole}}]{nagashima.etal.05}
{Nagashima}, M., {Lacey}, C.~G., {Baugh}, C.~M., {Frenk}, C.~S., \& {Cole}, S.
  2005, \mnras, 358, 1247


\bibitem[{{Pratt} {et~al.}(2007){Pratt}, {B{\"o}hringer}, {Croston}, 
{Arnaud},  {Borgani}, {Finoguenov},  {Temple}}]
        {pratt.etal.07} {Pratt}, G.~W., {B{\"o}hringer}, H., {Croston}, J.~H., 
{Arnaud}, M., {Borgani}, S., {Finoguenov}, A., {Temple}, R.~F. 2007, \aap,
        461, 71

\bibitem[{{Rasia} {et~al.}(2005){Rasia}, {Mazzotta}, {Borgani}, {Moscardini}, 
   {Dolag}, {Tormen}, {Diaferio}, \& {Murante}}]{rasia.etal.05}
{Rasia}, E., {Mazzotta}, P.,  {Borgani}, S.,  {Moscardini}, L., 
  {Dolag}, K., {Tormen}, G., {Diaferio}, A., \& {Murante}, G. 2005, \apjl,
  618,L1


\bibitem[{{Rasia} {et~al.}(2006){Rasia}, {Ettori}, {Moscardini}, {Mazzotta},
  {Borgani}, {Dolag}, {Tormen}, {Cheng}, \& {Diaferio}}]{rasia.etal.06}
{Rasia}, E., {Ettori}, S., {Moscardini}, L., {Mazzotta}, P., {Borgani}, S.,
  {Dolag}, K., {Tormen}, G., {Cheng}, L.~M., \& {Diaferio}, A. 2006, \mnras,
  369, 2013

\bibitem[{{Renzini} {et~al.}(1993){Renzini}, {Ciotti}, {D'Ercole}, \&
  {Pellegrini}}]{renzini.etal.93}
{Renzini}, A., {Ciotti}, L., {D'Ercole}, A., \& {Pellegrini}, S. 1993, \apj,
  419, 52

\bibitem[{{Romeo} {et~al.}(2006){Romeo}, {Sommer-Larsen}, {Portinari}, \&
  {Antonuccio-Delogu}}]{romeo.etal.06}
{Romeo}, A.~D., {Sommer-Larsen}, J., {Portinari}, L., \& {Antonuccio-Delogu},
  V. 2006, \mnras, 371, 548

\bibitem[{{Sakelliou} {et~al.}(2002){Sakelliou}, {Peterson}, {Tamura},
  {Paerels}, {Kaastra}, {Belsole}, {B{\"o}hringer}, {Branduardi-Raymont},
  {Ferrigno}, {den Herder}, {Kennea}, {Mushotzky}, {Vestrand}, \&
  {Worrall}}]{Sakelliou.etal.02}
{Sakelliou}, I., {Peterson}, J.~R., {Tamura}, T., {Paerels}, F.~B.~S.,
  {Kaastra}, J.~S., {Belsole}, E., {B{\"o}hringer}, H., {Branduardi-Raymont},
  G., {Ferrigno}, C., {den Herder}, J.~W., {Kennea}, J., {Mushotzky}, R.~F.,
  {Vestrand}, W.~T., \& {Worrall}, D.~M. 2002, \aap, 391, 903

\bibitem[{{Salpeter}(1955)}]{salpeter}
{Salpeter}, E.~E. 1955, \apj, 121, 161

\bibitem[{{Sanders} \& {Fabian}(2006)}]{sanders&fabian06}
{Sanders}, J.~S. \& {Fabian}, A.~C. 2006, \mnras, 371, 1483

\bibitem[{{Sanders} {et~al.}(2004){Sanders}, {Fabian}, {Allen}, \&
  {Schmidt}}]{sanders.etal.04}
{Sanders}, J.~S., {Fabian}, A.~C., {Allen}, S.~W., \& {Schmidt}, R.~W. 2004,
  \mnras, 349, 952

\bibitem[{{Saro} {et~al.}(2006){Saro}, {Borgani}, {Tornatore}, {Dolag},
  {Murante}, {Biviano}, {Calura}, \& {Charlot}}]{saro.etal.06}
{Saro}, A., {Borgani}, S., {Tornatore}, L., {Dolag}, K., {Murante}, G.,
  {Biviano}, A., {Calura}, F., \& {Charlot}, S. 2006, \mnras, 373, 397

\bibitem[{{Sato} {et~al.}(2007){Sato}, {Yamasaki}, {Ishida}, {Ishisaki},
  {Ohashi}, {Kawahara}, {Kitaguchi}, {Kawaharada}, {Kokubun}, {Makishima},
  {Ota}, {Nakazawa}, {Tamura}, {Matsushita}, {Kawano}, {Fukazawa}, \&
  {Hughes}}]{sato.etal.07}
{Sato}, K., {Yamasaki}, N.~Y., {Ishida}, M., {Ishisaki}, Y., {Ohashi}, T.,
  {Kawahara}, H., {Kitaguchi}, T., {Kawaharada}, M., {Kokubun}, M.,
  {Makishima}, K., {Ota}, N., {Nakazawa}, K., {Tamura}, T., {Matsushita}, K.,
  {Kawano}, N., {Fukazawa}, Y., \& {Hughes}, J.~P. 2007, preprint,
  astro-ph/0701328

\bibitem[{{Schindler} {et~al.}(2005){Schindler}, {Kapferer}, {Domainko},
  {Mair}, {van Kampen}, {Kronberger}, {Kimeswenger}, {Ruffert}, {Mangete}, \&
  {Breitschwerdt}}]{schindler.etal.05}
{Schindler}, S., {Kapferer}, W., {Domainko}, W., {Mair}, M., {van Kampen}, E.,
  {Kronberger}, T., {Kimeswenger}, S., {Ruffert}, M., {Mangete}, O., \&
  {Breitschwerdt}, D. 2005, \aap, 435, L25

\bibitem[{{Smith} {et~al.}(2001){Smith}, {Brickhouse}, {Liedahl}, \&
  {Raymond}}]{smith.etal.01}
{Smith}, R.~K., {Brickhouse}, N.~S., {Liedahl}, D.~A., \& {Raymond}, J.~C.
  2001, \apjl, 556, L91

\bibitem[{{Springel}(2005)}]{gadget2}
{Springel}, V. 2005, \mnras, 364, 1105

\bibitem[{{Springel} \& {Hernquist}(2002)}]{springel&hernquist02}
{Springel}, V. \& {Hernquist}, L. 2002, \mnras, 333, 649

\bibitem[{{Springel} \& {Hernquist}(2003)}]{springel&hernquist03}
---. 2003, \mnras, 339, 289

\bibitem[{{Sutherland} \& {Dopita}(1993)}]{sutherland&dopita}
{Sutherland}, R.~S. \& {Dopita}, M.~A. 1993, \apjs, 88, 253

\bibitem[{{Tamura} {et~al.}(2001){Tamura}, {Bleeker}, {Kaastra}, {Ferrigno}, \&
  {Molendi}}]{Tamura.etal.01}
{Tamura}, T., {Bleeker}, J.~A.~M., {Kaastra}, J.~S., {Ferrigno}, C., \&
  {Molendi}, S. 2001, \aap, 379, 107

\bibitem[{{Tamura} {et~al.}(2004){Tamura}, {Kaastra}, {den Herder}, {Bleeker},
  \& {Peterson}}]{Tamura.etal.04}
{Tamura}, T., {Kaastra}, J.~S., {den Herder}, J.~W.~A., {Bleeker}, J.~A.~M., \&
  {Peterson}, J.~R. 2004, \aap, 420, 135

\bibitem[{{Tormen} {et~al.}(1997){Tormen}, {Bouchet}, \&
  {White}}]{tormen.etal.97}
{Tormen}, G., {Bouchet}, F.~R., \& {White}, S.~D.~M. 1997, \mnras, 286, 865

\bibitem[{{Tornatore} {et~al.}(2007){Tornatore}, {Borgani}, {Dolag}, \&
  {Matteucci}}]{tornatore.etal.07}
{Tornatore}, L., {Borgani}, S., {Dolag}, K., \& {Matteucci}, F. 2007, \mnras

\bibitem[{{Tornatore} {et~al.}(2004){Tornatore}, {Borgani}, {Matteucci},
  {Recchi}, \& {Tozzi}}]{tornatore.etal.04}
{Tornatore}, L., {Borgani}, S., {Matteucci}, F., {Recchi}, S., \& {Tozzi}, P.
  2004, \mnras, 349, L19

\bibitem[{{Valdarnini}(2003)}]{valdarnini03}
{Valdarnini}, R. 2003, \mnras, 339, 1117

\bibitem[{{Vikhlinin}(2006)}]{vikh06}
{Vikhlinin}, A. 2006, \apj, 640, 710

\bibitem[{{Vikhlinin} {et~al.}(1998){Vikhlinin}, {McNamara}, {Forman}, {Jones},
  {Quintana}, \& {Hornstrup}}]{vikh.etal.98}
{Vikhlinin}, A., {McNamara}, B.~R., {Forman}, W., {Jones}, C., {Quintana}, H.,
  \& {Hornstrup}, A. 1998, \apj, 502, 558

\bibitem[{{Werner} {et~al.}(2006){Werner}, {B{\"o}hringer}, {Kaastra}, {de
  Plaa}, {Simionescu}, \& {Vink}}]{werner.etal.06a}
{Werner}, N., {B{\"o}hringer}, H., {Kaastra}, J.~S., {de Plaa}, J.,
  {Simionescu}, A., \& {Vink}, J. 2006, \aap, 459, 353

\bibitem[{{Yoshida} {et~al.}(2001){Yoshida}, {Sheth}, \&
  {Diaferio}}]{yoshida.etal.01}
{Yoshida}, N., {Sheth}, R.~K., \& {Diaferio}, A. 2001, \mnras, 328, 669

\end{thebibliography}
\end{document}